\documentstyle[aps,preprint,tighten,epsfig]{revtex}
\begin{document}
\draft
\def\etal{{\em et\ al.\/}}
\def\eg{{\em e.g.\/}}
\def\ie{{\em i.e.\/}, }
\def\beq{\begin{equation}}
\def\eeq{\end{equation}}
\def\beqarr{\begin{eqnarray}}
\def\eeqarr{\end{eqnarray}}
\def\ix{\int_{-L/2}^{L/2}dx\,}
\def\ixi{\int dx\,}
\def\ii{\int_{-\infty}^{\infty}}
\def\eps{\varepsilon}
\def\ov{\over}
\def\ol{\overline}
\def\sgn{{\rm sgn}}
\def\ph{\hat{\phi}}
\def\Pt{\tilde{\Psi}}
\def\tr{{\rm tr}}
\def\det{{\rm det}}
\def\la{\langle}
\def\ra{\rangle}
\def\8{\infty}
\def\one{\openone}
\def\lp{\lambda'}
\def\Zint{
  {\mathchoice
    {{\sf Z\hskip-0.45em{}Z}}
    {{\sf Z\hskip-0.44em{}Z}}
    {{\sf Z\hskip-0.34em{}Z}}
    {{\sf Z\hskip-0.35em{}Z}}
    }
  }
\def\del{\partial}
\def\delx{\partial_x}
\def\delxp{\partial_{x'}}
\def\delt{\partial_{\tau}}
\def\delT{\partial_t}
\def\XLL{{$\chi$LL}}
\def\gtlt{\,{\stackrel {\scriptscriptstyle >}{\scriptscriptstyle <}}\,}
\def\bh{{\hat \beta}}
\def\ph{{\hat \phi}}
\def\XSG{{$\chi$SG }}
\def\bJ{{\bf J}}
\def\Lt{L^{(1/2)}}
\def\Lh{L'^{(1/2)}}
\def\lh{{\hat \lambda}}
\def\nb{{\bar \nu}}
\def\tb{{\bar \tau}}
\def\ID{I}
\def\Ab#1{{\bf A}^{(#1)}}
\def\Bb#1{{\bf B}^{(#1)}}
\def\Cb#1{{\bf C}^{(#1)}}
\def\Af#1{\tilde{\bf A}^{(#1)}}
\def\Bf{\tilde{\bf B}}
\def\Cf#1{\tilde{\bf C}^{(#1)}}
\def\Rf#1{\tilde{\bf R}^{(#1)}}
\def\Gf#1{\tilde{\bf G}^{(#1)}}
\def\X#1{2\pi(x-v_{#1}t+i\epsilon_t)}
\def\XP#1{2\pi(x-x'-v_{#1}t+i\epsilon_t)}
\def\Tt{{\tilde T}}
\def\Lt{{\tilde L}}
\def\mat{\sf}
\def\Wt{{\widetilde W}}

\title{Edge Dynamics in Quantum Hall Bilayers II:\\ Exact Results with
Disorder and Parallel Fields}

\author{J. D. Naud$^1$, L. P. Pryadko$^2$, and S. L. Sondhi$^1$}

\address{
$^1$Department of Physics,
Princeton University,
Princeton, NJ 08544, USA}

\address{
$^2$School of Natural Sciences,
Institute for Advanced Study,
Olden Lane,
Princeton, NJ 08540}
\date{\today}
\maketitle

\begin{abstract}
We study edge dynamics in the presence of interlayer tunneling,
parallel magnetic field, and various types of disorder for two
infinite sequences of quantum Hall states in symmetric bilayers. These
sequences begin with the 110 and 331 Halperin states and include their
fractional descendants at lower filling factors; the former is easily
realized experimentally while the latter is a candidate for the
experimentally observed quantum Hall state at a total filling factor
of 1/2 in bilayers.  We discuss the experimentally interesting
observables that involve just one chiral edge of the sample and the
correlation functions needed for computing them.  We present several
methods for obtaining exact results in the presence of interactions
and disorder which rely on the chiral character of the system.  Of
particular interest are our results on the 331 state which suggest
that a time-resolved measurement at the edge can be used to
discriminate between the 331 and Pfaffian scenarios for the observed
quantum Hall state at filling factor 1/2 in realistic double-layer
systems.
\end{abstract}

\narrowtext
\section{Introduction}
\label{sec:intro}

The dynamics of the edge modes in quantum Hall systems has been a
subject of great interest for some years
\cite{halperin-edge,wen1,kanefisher}.  Its appeal is multifold.  The
low energy excitations of the ideal quantum Hall states that give rise
to the plateau in the Hall resistance exist only at the edges.  There
is a deep connection between the structure of the bulk ground states
and the ``universal content'' of the edge dynamics which is captured
mathematically in a relation between 2+1 dimensional Chern-Simons
theories and 1+1 dimensional conformal field theories
\cite{witten,moore}.  This connection in turn implies a non-trivial
charge dynamics at the edge which is now supported by experiments
\cite{chang,grayson}.  Finally, this connection allows the logic to be
turned around in deducing new quantum Hall states from an analysis of
possible conformal field theories \cite{moore,readrezayi}.

In this paper we investigate the edge dynamics of two infinite
sequences of quantum Hall states in statistically symmetric bilayer
systems in which we supplement the universal content by the inclusion
of interlayer electron tunneling, an additional magnetic field
parallel to the layers and, most importantly, disorder. The chief
interest of this problem is that interlayer tunneling is strongly
affected by the non-trivial charge dynamics and thus serves as an
``internal probe'' of the latter. In a previous publication
\cite{csgpaper}, henceforth {\bf I}, we had studied the problem of a
clean system that gives rise to a chiral version of the sine-Gordon
theory which is exactly soluble for the two infinite families of
states for which interlayer tunneling is not irrelevant: these are the
$mm'n$ Halperin states with $m=m'=n+1$ and $m=m'=n+2$. The second of
these families was shown to exhibit a remarkable {\it trifurcation} of
charged excitations on the edge with the appearance of two Majorana
fermions with dynamically generated velocities. In this paper we
consider the additional effect of disorder and a non-zero temperature
on the dynamics as well as the possibilities of modifying the
tunneling by means of an interlayer magnetic field or a gated transfer
of charge between the layers; the latter two procedures are
essentially equivalent as we will see below.

While the additional complications, especially the disorder, do not
allow a complete solution in the sense of finding distributions of
correlation functions in interlayer fields and at finite temperatures,
the chiral character of the dynamics still allows us to make
substantial progress in ways that should be of considerable interest
to readers with a background in the physics of interacting, disordered
systems. Consequently, we have included some amount of technical detail
in the paper. In order not to lose sight of the principal physical
results, especially those on the 331 state and the double layer
Pfaffian state that are experimentally testable, this introduction is
followed by a summary of the ``useful content'' of the paper.  Readers
primarily interested in the bottom line may wish to stop their perusal
at its end.

Before proceeding to that summary, a brief discussion of the
observables and the relevant correlation functions is in order. As we
will make more precise below, we study systems that possess one edge
mode per layer so their single layer analogs are the Laughlin states
at $\nu=1/m$ with $m$ odd. In those cases a {\it single} edge presents
three natural observables\cite{endnote1}.  The first is the ground
state expectation value of the edge current which can, in principle,
do interesting things if the flux through the bulk is
varied\cite{Ino}. The second is the tunneling density of states,
computed from the one-electron Green's function, and the last is the
edge mode velocity measured in a time resolved experiment done at the
edge, which enters the retarded density-density correlation
function. In real life, the first is not experimentally relevant,
while the third is not terribly interesting when there is one mode
that is unaffected by disorder or temperature. The first has been
experimentally investigated \cite{chang,grayson} to great effect if
not theoretical satisfaction, see Ref.~\cite{levitov} and references
therein.

One of our central contentions is that the collective mode structure
{\it is} interesting in double layer systems, even in the minimal cases
where there is only one edge mode per layer. This was already clear in
the clean cases considered previously, as in the trifurcation alluded to
above, and is also the case in the more involved and realistic
cases studied here. Consequently, the computation of retarded density
correlators central to time and layer resolved measurements at the
edges will be a central concern. In addition we will also compute one
electron Green's functions needed for tunneling measurements but they
will turn out to be essentially insensitive to the perturbations that we
consider. We will not compute edge currents, although our results on
partition functions in {\bf I} can be easily extended to do so in clean
systems, and extensions to the cases studied here are also feasible.

We should note that some of this work has technical connections to earlier
work on single layer systems \cite{wen2,kfp} with multiple edge modes
but the details are different and one of our sequences, inclusive of the
331 Halperin state relevant to experiments, has no analog in single layer
systems.

The outline of the paper is as follows.  We begin with a review of the
edge theory of clean bilayer systems in the presence of uniform
interlayer tunneling at the edge (Section~\ref{ss:reviewS}).  Next we
present our results for the effects of an interlayer magnetic field
(Section~\ref{ss:parallelfieldS}), disorder
(Section~\ref{ss:disorderS}), and a finite temperature
(Section~\ref{ss:finitetempS}).  The experimental consequences are
discussed in Section~\ref{ss:experimentS}, and the details of the
calculations are presented in Section~\ref{sec:details} and the
appendices.

\section{Summary of Results}
\label{sec:summary}

In this section we present a summary of our results.  Readers
interested in the details of the calculations will find them in
Section~\ref{sec:details}.  We begin with a review of the edge theory
of clean bilayer systems.  Next we consider the inclusion of a
parallel magnetic field, disorder, and a finite temperature.  We
conclude with an experimental proposal for an edge measurement that
could determine the bulk state responsible for the $\nu=1/2$ plateau
observed in bilayer systems.

\subsection{Review of Clean Bilayer Edge Theory}
\label{ss:reviewS}

\begin{figure}[htbp]
\begin{center}
\leavevmode
\epsfxsize=0.5\columnwidth
\epsfbox{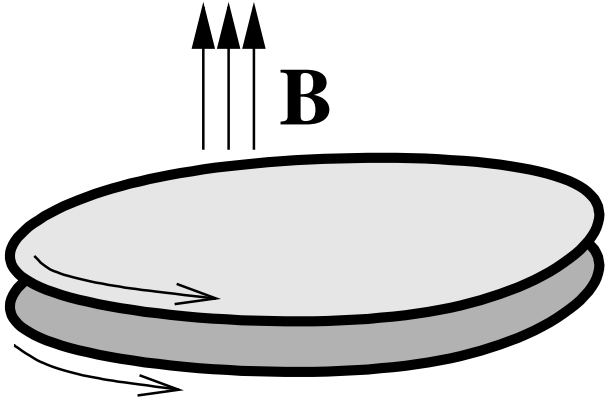}\vskip1mm
\caption{
The overall geometry of the bilayer quantum Hall system in a magnetic
field $B$ with edge modes in both layers propagating in the same direction. 
}
\label{fig:geo}
\end{center}
\end{figure}

The system under study consists of two parallel layers of 2DEGs in a
strong perpendicular magnetic field.  The geometry is sketched in 
Fig.~\ref{fig:geo}.  Specifically, we are interested
in the edge excitations of the Halperin states described by the $mmn$
wavefunction,

\widetext
\beq
\label{eq:wavefunc}
\Psi_{m,m,n}(\{z_{i\alpha}\})=\prod_{\alpha<\beta}(z_{1\alpha}-z_{1\beta})^m
(z_{2\alpha}-z_{2\beta})^{m}\prod_{\alpha,\beta}
(z_{1\alpha}-z_{2\beta})^n\,\, e^{-\sum_{i,\alpha}|z_{i \alpha}|^2/4},
\eeq
\narrowtext
where $z_{i \alpha}$ is the complex coordinate of electron $\alpha$ in
layer $i$.  The integer $m$ determines the correlations within the
layers and the integer $n$ specifies the interlayer correlations.
These states are incompressible and thus the gapless excitations of
the system are confined to the droplet edges, which have length $L$
and are parameterized by the coordinate $x$.

The edge theory contains two chiral Bose fields, a charged mode,
$\phi_{\rm c}$, and a neutral mode $\phi$.  We denote the velocities
of these modes by $v_{\rm c,n}$, respectively.  Excitations of the
charged mode correspond to charge being added to the edge from the
bulk, whereas excitations of the neutral mode correspond to 
a transfer of charge between the edges of the two layers.  We restrict
our discussion to states for which both edge modes move in the same
direction, the ``maximally chiral'' case, by requiring $m>n$.

In {\bf I} it was shown that in the presence of interlayer single
electron tunneling at the edge, the Hamiltonian of the edge theory
separates into a free chiral boson Hamiltonian for the charged mode
and a chiral sine-Gordon Hamiltonian ($\chi$SG) for the neutral mode.
The chiral sine-Gordon Hamiltonian depends on the tunneling strength,
denoted $\lambda$, and the parameter $\bh\equiv\sqrt{2(m-n)}$, which
sets the period of the interaction term.  Since the neutral mode
Hamiltonian depends only on $m-n$, the set of all maximally-chiral
bilayer states can be divided into sequences labeled by the value of
this difference.  In particular, we will concentrate on the 110
sequence, which contains all states with $m-n=1$, and the 331
sequence, composed of states with $m-n=2$.  The tunneling perturbation
is relevant, in a renormalization group (RG) sense, for the 110
sequence and marginal for the 331 sequence.

The 110 and 331 sequences were solved exactly in {\bf I}.
In particular the single-electron Green's function
\beq
\label{eq:prop}
{\cal G}_{ij}(t,x)\equiv -i\la T\,:\!\Psi_i(t,x)\!:\,
:\!\Psi^{\dag}_j(0,0)\!:\,\ra,
\eeq
and the two-point function of the density-fluctuation operator
\beq
\label{eq:rho_rho}
i{\cal D}_{ij}(t,x)\equiv \la T\rho_i(t,x)\rho_j(0,0)\ra
-\la\rho_i(t,x)\ra\la\rho_j(0,0)\ra,
\eeq
where $\Psi_i$ and $\rho_i$ are the electron annihilation and charge
density operators on the edge of layer $i$, respectively,
were computed exactly at zero temperature and $L\rightarrow \8$.  We
reproduce these results here, adopting a self-evident matrix notation.
For the 110 sequence we have
\widetext
\beqarr
{\cal G}(t,x)&=&
{\one\cos(\lambda x/v_{\rm n})+i\sigma^x \sin(\lambda x/v_{\rm n})\ov 
[\X {\rm c}]^{m-{1\ov 2}}\sqrt{\X {\rm n}}},
\label{eq:CSG_G110}\\
-i{\cal D}(t,x)&=&
{1\ov 2(2m-1)}{(\one+\sigma^x)\ov [\X {\rm c}]^2}
+{(\one-\sigma^x)\ov 2}{\cos(2\lambda x/v_{\rm n})\ov [\X {\rm n}]^2},
\label{eq:CSG_D110} 
\eeqarr
where $\one$ is the $2\times 2$ unit matrix and $\sigma^x$ is the
standard Pauli matrix.  For the 331 sequence we find
\beqarr
{\cal G}(t,x)&=&
{1\over \left[{\X {\rm c}}\right]^{m-1}}{1\ov 2}\left[{
{\one+\sigma^x\ov \X 1}+{\one-\sigma^x\ov \X 2}}\right],
\label{eq:CSG_G331}\\
-i{\cal D}(t,x)&=&
{1\ov 4(m-1)}{(\one+\sigma^x)\ov [\X {\rm c}]^2}
+{1\ov 4}{(\one-\sigma^x)\ov \X 1 \X 2},
\label{eq:CSG_D331} 
\eeqarr 
where we have defined the velocities $v_{1,2}\equiv v_{\rm n}\pm
\lambda/\pi$, and $\epsilon_t\equiv\sgn(t)a$, where $a$ is a short
distance cutoff.  Each of these functions contains a part arising from
the charged mode and a part from the neutral mode.  For the
single-electron Green's functions the contributions from each mode are
combined multiplicatively, while for the density-density correlation
function they are combined additively.

In addition to these time-ordered correlation functions, in later
sections we will be interested in the corresponding retarded
functions which govern physical response measurements at the edge.
The density response function is
\beq
\label{eq:CSG_DR110}
{\cal D}^R(t,x)=-{\theta(t)\ov 2\pi}\left[{
{(\one+\sigma^x)\ov 2(2m-1)}\delta'(x-v_{\rm c}t)
+{(\one-\sigma^x)\ov 2}\cos(2\lambda x/v_{\rm n})\delta'(x-v_{\rm
  n}t)}\right], 
\eeq
for the 110 sequence, and 
\beq
\label{eq:CSG_DR331} 
{\cal D}^R(t,x)=-{\theta(t)\ov 2\pi}\left[{
{(\one+\sigma^x)\ov 4(m-1)}\delta'(x-v_{\rm c}t)
+{(\one-\sigma^x)\ov 4(v_1-v_2)x}\left\{{v_2\delta(x-v_2t)
-v_1\delta(x-v_1t)}\right\}}\right],
\eeq
\narrowtext
for the 331 sequence, where the prime denotes
differentiation with respect to the argument.

We see that for the 110 sequence relevant tunneling produces spatial
oscillations in the correlation functions, while for the 331 sequence
marginal tunneling leads to two velocities ($v_{1,2}$) in the neutral
mode sector, and hence a total of {\em three} velocities for the
system as a whole.  Note that even at zero temperature and in the
absence of disorder the signal from the neutral mode in the
density-density response function decays with distance because of
tunneling.  In the following section we shall investigate how these
correlation functions are modified by various perturbations.

\subsection{Parallel Field}
\label{ss:parallelfieldS}

We first discuss the effects of an interlayer magnetic field.  If we
take the $z$-axis along the normal to the layers and recall that the
$x$-axis is along the edges, we consider an additional magnetic field
along the $y$-axis: ${\bf B}=B\hat{\bf y}$.  The edge Hamiltonian in
the presence of a parallel field depends on the parameter
$\Gamma\equiv v_{\rm n}Bd\bh/2$, where $d$ is the layer separation.
We remark that the effect of the interplane magnetic field considered
here is distinct from the simple decrease in $\lambda$ caused by the
reduction of the interlayer tunneling matrix element.  As noted by
Chalker and Sondhi, these effects can be distinguished by studying
large aspect-ratio samples\cite{chalker-sondhi}.
The results here also apply to the case where we introduce an
electric potential difference between the layers instead of an
interplane magnetic field.  

\subsubsection{110 Sequence}

For the states in the 110 sequence we find that the
spectrum of the edge theory in the presence of an interlayer magnetic
field can be obtained from the spectrum with zero interlayer field
via the replacement
\beq
\label{eq:lambdaprimeS}
\lambda\mapsto \lp\equiv\sqrt{\lambda^2 +\Gamma^2/2},
\eeq
in particular we see that the number of velocities is unchanged.  The
two-point function of the density-fluctuation operator
(\ref{eq:rho_rho}) is
\beqarr
\lefteqn{-i{\cal D}(t,x)={1\ov 2(2m-1)}{(\one+\sigma^x)\ov [\X {\rm
      c}]^2}+{1\ov 2 (\lp)^2}} 
\nonumber\\&&{}\times{(\one-\sigma^x)\ov [\X {\rm n}]^2}
\left\{{{\Gamma^2\ov 2}+\lambda^2\cos\left({{2\lp\,x
\ov v_{\rm n}}}\right)}\right\}.\label{eq:rho_rho_w/BS}
\eeqarr
Note that in the absence of tunneling ($\lambda=0$), the correlation
function is unaffected by the parallel magnetic field, \ie it is
independent of $\Gamma$, as expected.  For non-zero tunneling, the
addition of an interlayer magnetic field increases the frequency and
decreases the amplitude of the spatial oscillations in the
density-density correlation function.  One can show that the effect of
the parallel field on the single-electron Green's function is similar
to its effect on the density two-point function, see Eq.~(\ref{eq:G_B}).

\subsubsection{331 Sequence}

\begin{figure}[htbp]
\begin{center}
\leavevmode
\epsfxsize=0.6\columnwidth
\epsfbox{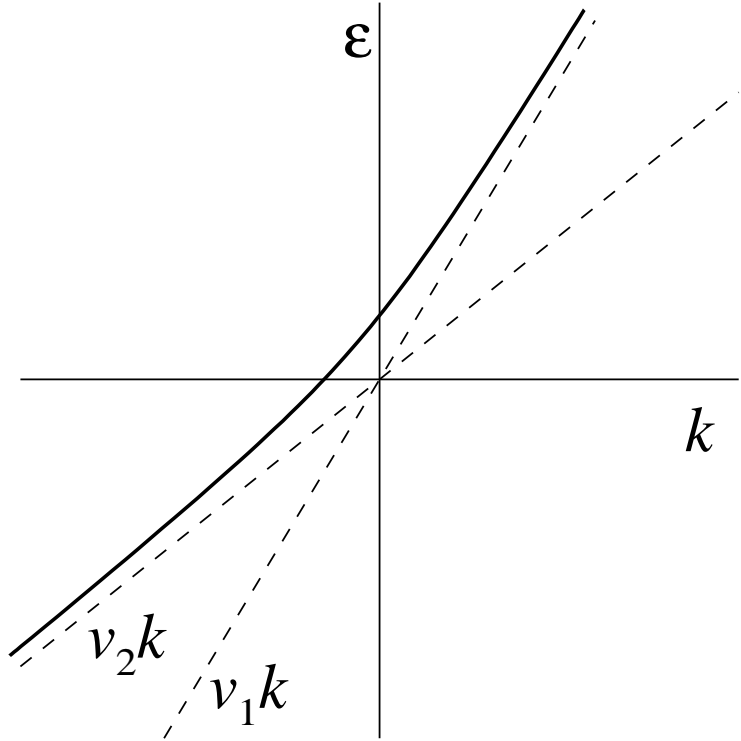}\vskip1mm
\caption{The solid line is the dispersion $\eps(k)$ plotted for $\Gamma>0$.
The dotted line is $\eps=v_2k$ and the dashed line is $\eps=v_1k$.
Note that $\eps(k)$ asymptotes $v_1k$ as $k\rightarrow\8$ and $v_2k$
as $k\rightarrow-\8$.}
\label{fig:disp}
\end{center}
\end{figure}

For the states in the 331 sequence we find that 
in the presence of an interlayer magnetic field the spectrum of the
neutral mode portion of the edge theory is
\beqarr
\label{eq:H/w/B/diagS}
{\cal H}_B&=&\sum_k\eps(k)\,:\!a_k^{\dag}a_k\!:,
\quad{\rm where}\quad\\
\nonumber
 \eps(k)&\equiv& v_{\rm n}k+\sgn(\Gamma)
\sqrt{(\lambda k/\pi)^2+\Gamma^2},
\eeqarr
where the $a_k$ are canonical Fermi operators.  Note that in the limit
of vanishing parallel magnetic field, $\Gamma\rightarrow 0^+$, the
energy dispersion (for $\lambda >0$) becomes $\eps(k)=(v_{\rm
n}+\lambda/\pi)k=v_1k$ for $k>0$ and $\eps(k)=(v_{\rm
n}-\lambda/\pi)k=v_2k$ for $k<0$, which is the spectrum of two
right-moving Majorana fermions with split velocities.  (In the limit
$\Gamma\rightarrow 0^-$, or for $\lambda <0$, we get a similar
dispersion with $v_1$ and $v_2$ interchanged, but this does not alter
the excitation spectrum of the Hamiltonian).  For any non-zero
interplane field we find that the dispersion develops some curvature,
which corresponds to the two Majorana species being mixed at distances
large compared with $\lambda/\pi\Gamma$.  The dispersion is sketched
in Fig.~\ref{fig:disp}.

The correlation functions for the 331 sequence with a parallel field
can be reduced to quadrature.  If we define the quantities $
\tau\equiv\sgn(t)(v_{\rm n}X-t)$,
\beq
X\equiv{x\ov v_1v_2},\qquad
\kappa(\omega)\equiv\sqrt{v_1v_2\Gamma^2+(\lambda/\pi)^2\omega^2},
\label{eq:quantS}
\eeq
and the function
\beq
\label{eq:PS}
P(\tau,X)\equiv\int_0^{\8}{d\omega\ov 2\pi}\,e^{i\omega\tau}
{\sin(\kappa(\omega)X)\ov \kappa(\omega)},
\eeq
then the single-electron Green's function and the density-density
correlation function can be expressed in terms of $P(\tau,X)$ and
its derivatives.  We find
\widetext
\beqarr
{\cal G}(t,x)&=&{-i\sgn(t)\ov v_1v_2 \left[{\X {\rm c}}\right]^{m-1}}
\left[{\left({v_{\rm n}\one -{\lambda\ov\pi}\sigma^x}\right)P_X(\tau,X)
}\right.\nonumber\\
&&{}+\sgn(t) \left({{\lambda\ov\pi}\one-v_{\rm
      n}\sigma^x}\right){\lambda\ov\pi} 
P_{\tau}(\tau,X)-iv_1v_2\Gamma
\sigma^zP(\tau,X)\biggr],\label{eq:G_psi_BS}\\
-i{\cal D}(t,x)&=&{1\ov 4(m-1)}{(\one+\sigma^x)\ov [\X {\rm c}]^2}
-{(\one-\sigma^x)\ov 4v_1v_2}\left[{P_X^2-{\lambda^2\ov \pi^2}P_{\tau}^2
+v_1v_2\Gamma^2P^2}\right],
\label{eq:DDw/B,331,2S}
\eeqarr
\narrowtext\noindent
where the subscripts on $P$ denote partial differentiation.

\begin{figure}[htbp]
\begin{center}
\leavevmode
\epsfxsize=0.7\columnwidth
\epsfbox{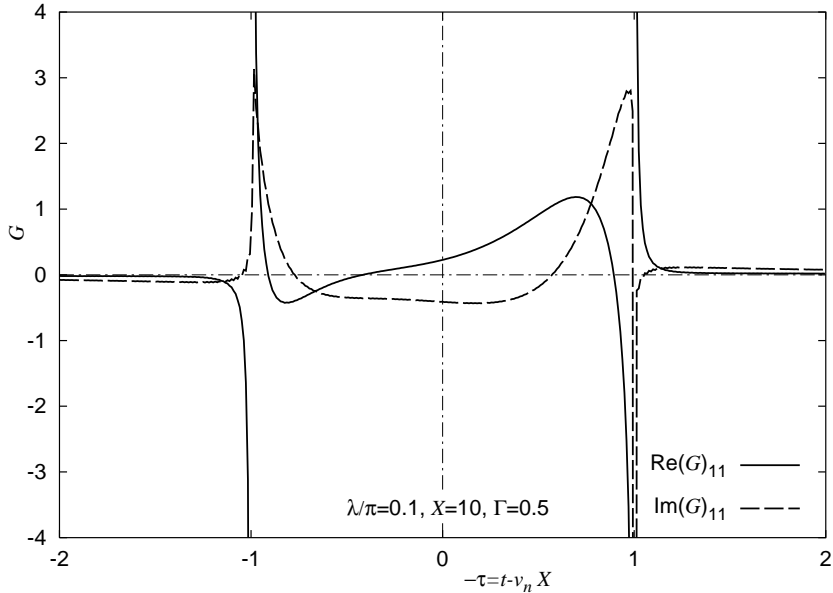}\vskip1mm
\caption{
The real (solid line) and imaginary (dashed line) parts of the neutral
mode factor in ${\cal G}_{11}$ (\protect\ref{eq:G_psi_B}) plotted as a
function of $-\tau$ for $v_{\rm n}=1$, $\lambda/\pi=0.1$, $X=10$, and
$\Gamma=0.5$.
}
\label{fig:bpla11}
\end{center}
\end{figure}

\begin{figure}[htbp]
\begin{center}
\leavevmode
\epsfxsize=0.7\columnwidth
\epsfbox{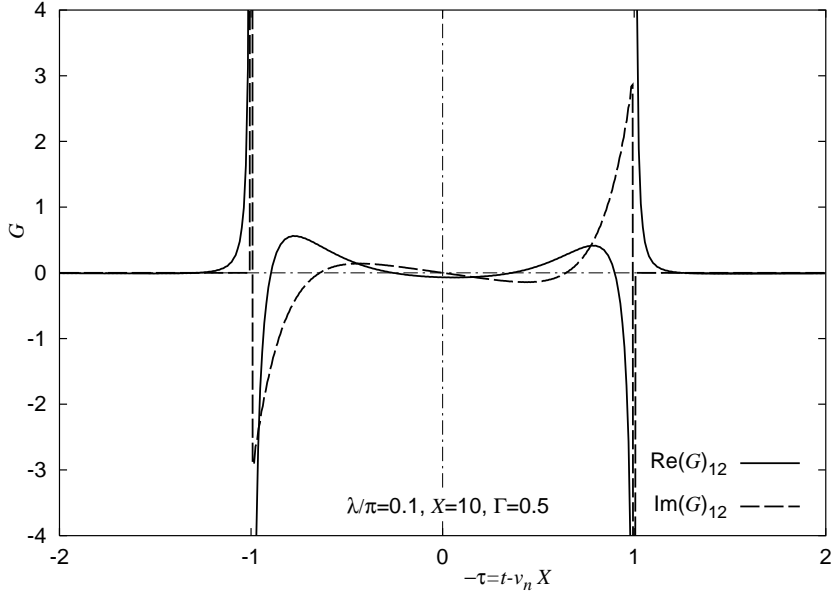}\vskip1mm
\caption{
The real (solid line) and imaginary (dashed line) parts of the neutral
mode factor in ${\cal G}_{12}$ (\protect\ref{eq:G_psi_B}) plotted as a
function of $-\tau$ for $v_{\rm n}=1$, $\lambda/\pi=0.1$, $X=10$, and
$\Gamma=0.5$.
}
\label{fig:bpla12}
\end{center}
\end{figure}

In Fig.~\ref{fig:bpla11} we plot the real and imaginary parts of the
neutral mode part of ${\cal G}_{11}$ at fixed $X=10$ as a function of
$-\tau$ for $v_{\rm n}=1$, $\Gamma=0.5$, and $\lambda/\pi=0.1$.  The
corresponding plot for the neutral mode part of ${\cal G}_{12}$ is
given in Fig.~\ref{fig:bpla12}.  The singularities visible in these
plots occur at the points $t=x/v_1$ and $t=x/v_2$, corresponding to
propagation at the Majorana velocities.  We see that the parallel
field does not destroy the velocity-split structure of the Green's
function, which is somewhat remarkable.  The parallel field is an RG
relevant perturbation which modifies the low-energy, long-wavelength
properties of the system, see Fig.~\ref{fig:disp}.  Nevertheless, the
singularities at $v_1$ and $v_2$ are completely unchanged, see
Eq.~(\ref{eq:gR_wB}) below, since they arise from integrations over all
frequencies.
In Appendix~\ref{sec:random-tunneling} we
will discuss the opposite case: an RG irrelevant perturbation which
does modify some features of the Green's function.  
At times
corresponding to propagation velocities between $v_1$ and $v_2$ we
find oscillatory behavior not present at $\Gamma=0$.

\subsection{Disorder}
\label{ss:disorderS}

We now consider the effects of disorder on the bilayer system.  Our
primary interest is how the novel features of the correlation
functions in the presence of tunneling, \ie spatial oscillations in
the 110 sequence and the splitting of the neutral-mode Majorana
velocities in the 331 sequence, are modified by disorder.  Note that
the quantities of physical interest are the correlation functions in a
typical realization of disorder, whereas the readily calculable
quantities are disorder-averaged correlation functions.  The typical
and average quantities may have very different behavior, and we will
discuss such differences at various points.

In addition to the possibility of disorder in the tunneling amplitude,
$\lambda$, we consider random scalar potentials, $\xi_i(x)$, which
couple to the edge charge densities in each layer.  From a
perturbative RG analysis we find that disorder in $\lambda$ is
relevant for the 110 sequence and irrelevant for the 331 sequence,
while the random scalar potentials, $\xi_i(x)$, are relevant for both
sequences.

\subsubsection{110 Sequence}

We consider a tunneling amplitude which has mean $\lambda$ and
variance $\Delta_{\lambda}$, and a disordered scalar potential with
zero mean and variance $\Delta_{\xi}$.  We find that the
disorder-averaged, retarded density response function is
\widetext
\beqarr
\lefteqn{\ol{{\cal D}^R}(t,x,x')=-{\theta(t)\ov 2\pi}\Biggl[
{(\one+\sigma^x)\ov 2(2m-1)}\delta'(x-x'-v_{\rm c}t)
+{(\one-\sigma^x)\ov 2}\delta'(x-x'-v_{\rm n}t)}&&
\nonumber \\
&&\left.{{}\times e^{-{2|x-x'|(\Delta_{\lambda}+\Delta_{\xi}/4)/v_{\rm n}^2}}
\left[{\cos\left({2|x-x'|\tilde{\lambda}\ov v_{\rm n}}\right)+
{\Delta_{\xi}\ov4v_{\rm n}\tilde{\lambda}}
\sin\left({2|x-x'|\tilde{\lambda}\ov v_{\rm n}}\right)}\right]}\right],
\label{eq:DR110_fint2S} 
\eeqarr
\narrowtext
where
\beq
\label{eq:shiftlambdaS}
\tilde{\lambda}\equiv
\sqrt{\lambda^2-\left({\Delta_{\xi}\ov 4v_{\rm n}}\right)^2}.
\eeq
The disorder in the tunneling amplitude ($\Delta_{\lambda}$)
produces an exponential decay with distance in the neutral mode part
of the disorder-averaged density response function.  The random scalar
potential ($\Delta_{\xi}$) has a similar effect, and in addition it
produces a shift in the frequency of the spatial oscillations
(\ref{eq:shiftlambdaS}).  Using the fact that the neutral-mode part of
${\cal D}$ in a given sample can be expressed in terms of products of
single-particle Green's functions, whose absolute squares are
long-ranged (\ie algebraically decaying) we can conclude that in a given
sample the density response function has the structure of the
disorder-averaged quantity (\ref{eq:DR110_fint2S}), without the
exponential decay in space of the neutral mode piece.

\subsubsection{331 Sequence}

Above we remarked that for the 331 sequence only disorder in the
scalar potential terms was a non-irrelevant perturbation.  We
therefore consider only a disordered scalar potential with zero mean
and variance $\Delta_{\xi}$.  We find that the disorder-averaged,
retarded density response function at $T=0$ is:
\widetext
\beq
\label{eq:DRFull}
\ol{{\cal D}^R}(t,x,x')=-{\theta(t)\ov 2\pi}
{\one +\sigma^x\ov 4(m-1)}\delta'(x-x'-v_{\rm c}t)
+{\one-\sigma^x\ov 4}\ol{D^R}(t,x,x'),
\eeq
where the neutral mode contribution is
\beqarr
\ol{D^R}(t,x,0)&=&{\theta(t)\ov 2\pi}e^{-\Delta_{\xi}x/v_1v_2}
\left[{{1\ov(v_1-v_2)x}\left\{{v_1\delta(x-v_1t)-v_2\delta(x-v_2t)}\right\}
+{\Delta_{\xi}\ov 2v_1v_2}\theta(z)}\right.\nonumber\\
&&\times\left.{{\cal P}\left({1\ov t_0-t}\right)
\left({{x/v_1v_2\ov\sqrt{z}}
I_1\left[{{\Delta_{\xi}\ov\lambda/\pi}\sqrt{z}}\right]
+{\pi\ov \lambda}
I_0\left[{{\Delta_{\xi}\ov\lambda/\pi}\sqrt{z}}\right] 
}\right)}\right].\label{eq:DisAvgDR331S}
\eeqarr
\narrowtext
Here $z\equiv (t-x/v_1)(x/v_2-t)$, $I_n$ are Bessel functions of
imaginary argument, and $t_0\equiv (x/v_1+x/v_2)/2=v_{\rm n}x/v_1v_2$
is the mean arrival time.

Comparing this expression to the result for the neutral mode in a
clean system, the second term in Eq.~(\ref{eq:CSG_DR331}), we find
that the delta-function peaks at the velocities $v_1$ and $v_2$ remain
sharp in the presence of disorder, but their amplitudes acquire an
additional exponential decay.  In the disorder-averaged result,
(\ref{eq:DisAvgDR331S}), there is also a signal centered around the
mean arrival time $t_0$.  If we write $\tau=t_0-t$, then in the limit
of large distances: $x/v_1v_2\gg 1/\Delta_{\xi}$, for times near the
mean arrival time, $(\pi/\lambda)\tau \ll x/v_1v_2$, the central
signal is asymptotically
\beq
\label{eq:largex_DR}
\ol{D^R}(t,x,0)\approx \sqrt{\Delta_{\xi}\ov x}{\cal P}
\left({1\ov\tau}\right)
\exp\left[{-{\Delta_{\xi}v_1v_2\ov 2x(\lambda/\pi)^2}\tau^2}\right].
\eeq
This asymptotic form is similar to a result obtained by Wen for the
case of two $\nu=1$ edges with unequal velocities\cite{wen2}.  The
reason for this similarity is that both problems are formally
equivalent to a spin in a random magnetic field which undergoes
diffusion on the SU(2) group manifold, see
Appendix~\ref{sec:331-via-spin}.  The term in Eq.~(\ref{eq:largex_DR})
decays algebraically with distance.  Therefore, while the signal in
$\ol{D^R}$ at the extremal velocities ($v_{1,2}$) is exponentially
suppressed by the disorder, there is an additional signal with
velocity $v_1v_2/v_{\rm n}$ which only falls off algebraically.

To determine the behavior of $D^R$ in a given realization of disorder
we have adopted several approaches.  First, as in our analysis of the
110 sequence, we can use the fact that $D^R$ can be expressed in terms
of single-particle Green's functions whose second moments we can
evaluate.  Second, we have found that $D^R(t,x,x')$ exhibits an exact
antisymmetry about the point $t=t_0$ in each realization of disorder:
\beq
\label{eq:sym_DRS}
D^R(t_0+t,x,x')=-D^R(t_0-t,x,x').
\eeq
This is an interesting result because it is an exact dynamical
symmetry (\ie it is a relation between correlation functions at
different times) in a system with an arbitrary potential $\xi(x)$.
Finally, we have calculated the behavior of the correlation functions
for two simple potentials, the case of a uniform potential
$\xi(x)={\rm const}$, and the case of isolated delta-function
impurities $\xi(x)=\sum_m q_m\delta(x-y_m)$.  All of these
results, discussed in detail in Section~\ref{ss:disorder}, lead us to
the conclusion that $D^R(t,x,x')$ in a typical configuration is given
by an expression similar to Eq.~(\ref{eq:DisAvgDR331S}) for
$\ol{D^R}(t,x,x')$,  with the principal value 
factor replaced by a function $f(t,x,x')$, which is a
rapidly fluctuating function of time, antisymmetric about the point
$t=t_0$, and whose amplitude grows as $t$ approaches $t_0$.
These conclusions about the behavior of $D^R$ in a given sample
have been verified by numerical simulations which are discussed in
Section~\ref{ss:experimentS}.

\subsection{Finite Temperature Effects}
\label{ss:finitetempS}

We now briefly consider the effects of a finite temperature
$T=1/\beta$.  For a single chiral edge mode we know that at zero
temperature ${\cal D}^R(t,x)\propto \theta(t)\delta'(x-vt)$.  A
straightforward calculation shows that this form is actually
temperature independent.

Recall that for the 110 sequence the retarded density-density
correlation function is a sum of terms of the form
$\theta(t)\delta'(x-vt)$ multiplied by a function independent of $t$.
This can be seen for the clean system in Eq.~(\ref{eq:CSG_DR110}), and
for the disordered system in Eq.~(\ref{eq:DR110_fint2S}).  We can
therefore conclude that ${\cal D}^R$ for the 110 sequence is
temperature independent even in the presence of a non-zero scalar
potential.

The situation is different for the 331 sequence.  While the term in
${\cal D}^R$ from the charged mode is temperature independent for the
reasons given above, the neutral mode term is not.  At a finite
temperature one finds for the neutral mode in a clean system:
\widetext
\beq
\label{eq:DR331_fint}
D^R(t,x)=-{\theta(t)\ov 2\beta v_1v_2}
{1\ov\sinh\left[{\pi(v_1-v_2)x/\beta v_1v_2}\right]}
\left[{v_2\delta(x-v_2t)-v_1\delta(x-v_1t)}\right].
\eeq
Comparing this with the zero temperature result, the second term in
Eq.~(\ref{eq:CSG_DR331}), we find that the neutral mode term, which
decays as $1/x$ at $T=0$, decays exponentially at $T>0$.

In the presence of disorder one finds that the finite-temperature form of
$\ol{D^R}$ is 
\beqarr
\lefteqn{\ol{D^R}(t,x,0)={\theta(t)\ov 2\pi}e^{-\Delta_{\xi}x/v_1v_2}
\left[{{(\pi/\beta v_1v_2)\ov\sinh[\pi(v_1-v_2)x/\beta v_1v_2]}
\left\{{v_1\delta(x-v_1t)-v_2\delta(x-v_2t)}\right\}
}\right.}&&\nonumber \\
&&{}+\Delta_{\xi}\theta(z)
{\cal P}\left({(\pi/\beta v_1v_2)\ov\sinh[2\pi(t_0-t)/\beta]}\right)
\left.{\left({{x/v_1v_2\ov\sqrt{z}}
I_1\left[{{\Delta_{\xi}\ov\lambda/\pi}\sqrt{z}}\right]
+{\pi\ov \lambda}
I_0\left[{{\Delta_{\xi}\ov\lambda/\pi}\sqrt{z}}\right] 
}\right)}\right].\label{eq:DR331_dirt_fint}
\eeqarr
\narrowtext
This result was obtained using the formalism given in
Appendix~\ref{sec:331-via-spin}.

Comparing this with the zero temperature result in
Eq.~(\ref{eq:DisAvgDR331S}) we see that the amplitudes of the
delta-functions at the extremal velocities acquire an additional
exponential decay because of the finite temperature.  However, the
replacement 
\beq
\label{eq:PV_fint}
{\cal P}\left({1\ov t_0-t}\right)\mapsto 
{\cal P}\left({(2\pi/\beta)\ov\sinh[2\pi(t_0-t)/\beta]}\right)
\eeq
indicates that the structure in $\ol{D^R}$ centered on the mean
arrival time {\em sharpens} at a finite temperature.

\subsection{Experimental Ramifications}
\label{ss:experimentS}

One of the primary results of the previous sections is the unusual
structure of the retarded density-density correlation function for the
331 sequence.  The first state in the 331 sequence, the state 331
itself, has a filling factor of 1/4 per layer, for a total filling
factor of 1/2.  A plateau in the Hall conductance has been
experimentally observed at $\nu=1/2$ in bilayer
systems\cite{eisenstein}.  Another candidate state which has been
proposed to explain this plateau is the Pfaffian state\cite{greiter}.
Standard experimental probes of the edge states, such as the
non-linear $I$--$V$ characteristic, cannot be used to distinguish the
331 from the Pfaffian state since both states give the same power law
exponent\cite{imura-ino}.  In this section, we argue that the retarded
density-density correlation function of the Pfaffian is sufficiently
different from that predicted for the 331 state that, even in the
presence of a finite temperature and disorder, a measurement of this
correlation function at the edge could distinguish between these two
bulk states.

For the Pfaffian edge theory we find that the retarded density
response function at a finite temperature and in the presence of
disorder is
\beq
\label{eq:PfDRS}
{\cal D}^R_{\rm Pf}(t,x)=-{1\ov 4\pi}\theta(t)\delta'(x-v_{\varphi}t).
\eeq
We see that there is only a single velocity present.

Recall from Section~\ref{ss:reviewS}, Eq.~(\ref{eq:CSG_DR331}), that
for the 331 edge there are three velocities present in the clean
system at zero temperature, one for the charged mode ($v_{\rm c}$) and
two for the neutral mode ($v_{1,2}$), whose splitting is due to
tunneling.  The signal at the two neutral mode velocities decays as
$1/x$ at $T=0$ in the clean system.  In Section~\ref{ss:disorderS} we
saw that a disordered scalar potential suppresses the signal at the
extremal velocities by a factor which decays exponentially with
distance.  However, we found that there is a broad signal in the
neutral mode, centered on a velocity distinct from the charged mode
velocity, which decays only algebraically.  In
Section~\ref{ss:finitetempS} we saw that a finite temperature
actually sharpens the signal centered on the mean arrival time.  From
the results and discussion in Sections~\ref{ss:disorderS} and
\ref{ss:finitetempS}, we expect that for the 331 edge in a given
realization of disorder, the structure of the retarded
finite-temperature density-density correlation function is: 
\beqarr
\label{eq:331calDR}
\lefteqn{{\cal D}^R_{331}(t,x,x')\approx
\theta(t)(\one+\sigma^x)\delta'(x-v_{\rm c}t) }&&\nonumber\\ 
&\quad&\quad+\theta(t)(\one-\sigma^x){f(t,x,x',\beta) \ov \sqrt{\ell(x-x')}}
\exp\left[{-{(t-t_0)^2\ov \ell(x-x')}}\right],
\eeqarr
where the function $f(t,x,x',\beta)$ is the finite-temperature version
of the function introduced at the end of Section~\ref{ss:disorderS},
and $\ell$ is a length scale set by the potential.  We expect
$f(t,x,x',\beta)$ to have the same properties as the zero-temperature
function, and to have its support more strongly concentrated near
$t=t_0$ as the temperature increases.

\begin{figure}[htbp]
\begin{center}
\leavevmode
\epsfxsize=0.8\columnwidth
\epsfbox{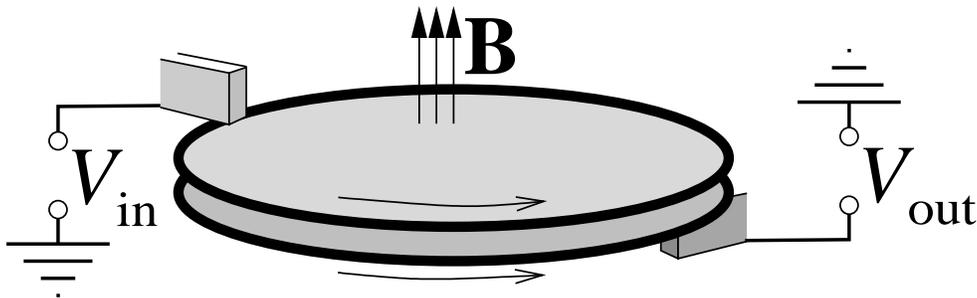}\vskip1mm
\caption{
The experimental geometry showing the quantum Hall bilayer in a
magnetic field $B$ with two spatially separated contacts.  A density
disturbance is produced at one electrode ($V_{\rm in}$) and measured
at the other electrode ($V_{\rm out}$).
}
\label{fig:exp}
\end{center}
\end{figure}

The experiment we propose involves creating a density disturbance at
one point along the edge of the bilayer system and measuring the
signal some distance downstream.  The experimental geometry is
sketched in Fig.~\ref{fig:exp}.  Basic linear response theory states
that if the density disturbance is produced in layer $j$ via an
external potential $V_{\rm ex}(t,x)$, and measured in layer $i$ then
the signal is
\beq
\label{eq:lin_resp}
\la\rho_i(t,x)\ra_{\rm ex}=\int dt'dx'\,{\cal D}^R_{ij}(t-t',x,x')
V_{\rm ex}(t',x').
\eeq
If the external potential is turned on at a point, \ie $V_{\rm
ex}(t',x')= \delta(x')\theta(t')$, then the measured signal is
\beq
\label{eq:lin_resp2}
\la\rho_i(t,x)\ra_{\rm ex}=\int_{-\8}^{t}dt'\,{\cal D}^R_{ij}(t',x,0).
\eeq

\begin{figure}[htbp]
  \begin{center}
  \leavevmode 
  \epsfxsize=0.7\columnwidth
  \epsfbox{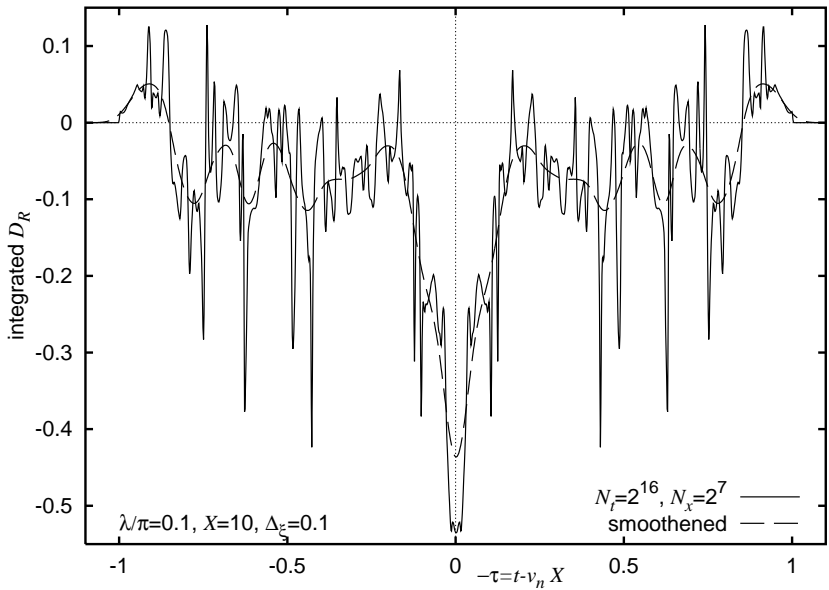}
  \vskip1mm 
  \caption{ 
    The neutral mode contribution to the integrated ${\cal D}^R_{12}$
    (\protect\ref{eq:lin_resp2}) at $T=0$ for $v_{\rm n}=1$,
    $\lambda/\pi=0.1$, $X=10$, and $\Delta_{\xi}=0.1$.  The horizontal
    axis is time measured from the mean arrival time.  The solid line
    is for a given realization of disorder, and the dashed line
    assumes a measurement with a finite time resolution.}
\label{fig:distrace} 
\end{center}
\end{figure}

\begin{figure}[htbp]
  \begin{center}
  \leavevmode 
  \epsfxsize=0.7\columnwidth
  \epsfbox{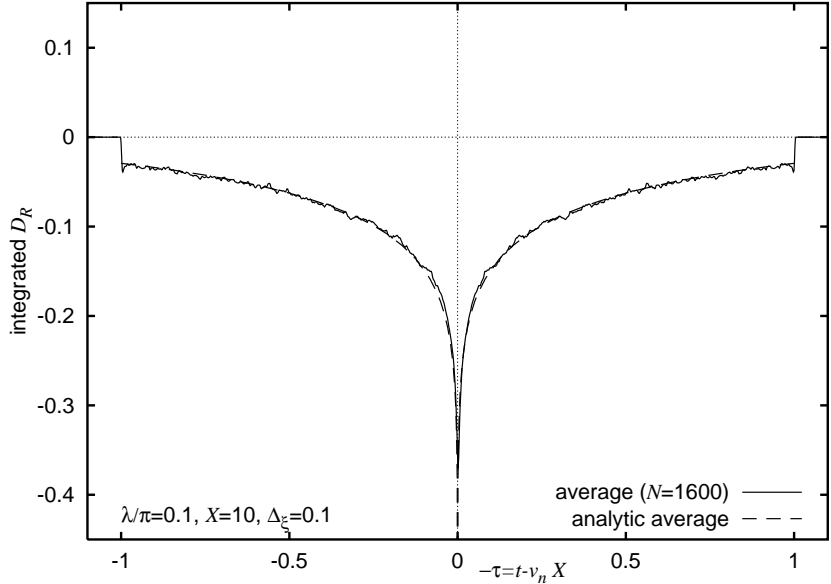}
  \vskip1mm 
  \caption{ 
    The neutral mode contribution to the integrated ${\cal D}^R_{12}$
    (\protect\ref{eq:lin_resp2}) at $T=0$ for $v_{\rm n}=1$,
    $\lambda/\pi=0.1$, $X=10$, and $\Delta_{\xi}=0.1$.  The solid line
    is the numerical average over 1600 impurity configurations and the
    dashed line is the analytical result.}
    \label{fig:distrace2} 
\end{center}
\end{figure}

For the Pfaffian state one would see a single sharp signal, see
Eq.~(\ref{eq:PfDRS}).  In contrast, for the 331 state, in addition to
a sharp signal from the charged mode there would be a second signal
{}from the neural mode.  To illustrate the neutral mode signal one would
expect in this case we have performed numerical simulations.  A
typical trace, computed at zero temperature for $v_{\rm n}=1$,
$\lambda/\pi=0.1$, $X=10$, $\Delta_{\xi}=0.1$ with stepwise-constant
disorder potential with $N_x=2^7$ values is shown in
Fig.~\ref{fig:distrace} with a solid line.  As expected, it is a
rapidly fluctuating function, but it exhibits an exact symmetry about
the mean arrival time.  This symmetry follows from the antisymmetry of
$D^R$ (\ref{eq:sym_DRS}) and the time integration
(\ref{eq:lin_resp2}).  Although the signal is very noisy, a
measurement with finite resolution (dashed line) produces a curve
which does not average to zero.  The amplitude of the smoothened
signal is maximal at the mean arrival time as was predicted in
Section~\ref{ss:disorderS}.  In Fig.~\ref{fig:distrace2} we show the
result of numerically averaging over 1600 impurity configurations
(solid line), as well as the analytic average (dashed line) evaluated
using Eq.~(\ref{eq:DisAvgDR331S}).  The two curves are in excellent
agreement.  The finite width of the neutral mode signal is a novel
feature of the 331 state; in the 110 sequence the neutral mode
propagates with only one velocity.

\begin{figure}[htbp]
  \begin{center}
  \leavevmode 
  \epsfxsize=0.7\columnwidth
  \epsfbox{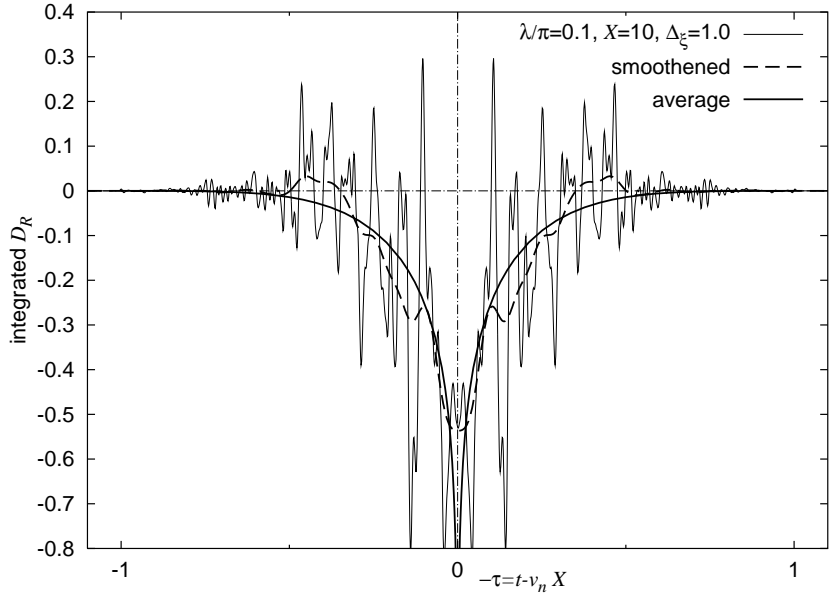}
  \vskip1mm 
  \caption{ 
    The neutral mode contribution to the integrated ${\cal D}^R_{12}$
    (\protect\ref{eq:lin_resp2}) at $T=0$ for $v_{\rm n}=1$,
    $\lambda/\pi=0.1$, $X=10$, and $\Delta_{\xi}=1.0$.  The thin solid line
    is for a given realization of disorder, the dashed line assumes a
    measurement with a finite time resolution, and the thick solid line
    is the analytical average.}
\label{fig:bigdirt} 
\end{center}
\end{figure}

\begin{figure}[htbp]
  \begin{center}
  \leavevmode 
  \epsfxsize=0.7\columnwidth
  \epsfbox{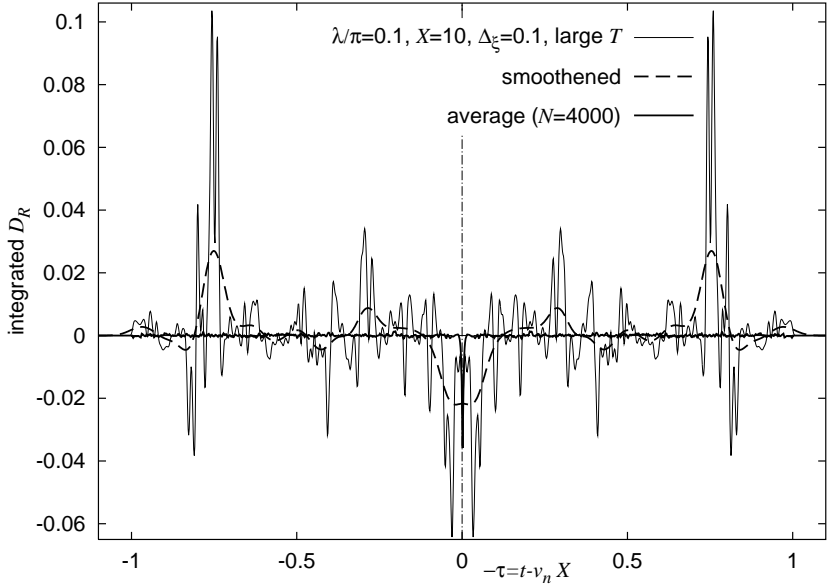}
  \vskip1mm 
  \caption{ 
    The neutral mode contribution to the integrated ${\cal D}^R_{12}$
    (\protect\ref{eq:lin_resp2}) at $T/\Delta_{\xi}=2000$ for $v_{\rm
    n}=1$, $\lambda/\pi=0.1$, $X=10$, and $\Delta_{\xi}=0.1$.  The thin
    solid line is for a given realization of disorder, the dashed line
    assumes a measurement with a finite time resolution, and the thick
    solid line is the result of averaging over 4000 impurity
    configurations}
\label{fig:finitetemp} 
\end{center}
\end{figure}

In Fig.~\ref{fig:bigdirt} we show a similar trace, but with the
disorder stronger by an order of magnitude, $\Delta_{\xi}=1.0$.  Note
that the amplitude of the signal near the extremal arrival times
$\tau=\pm 1$ is suppressed relative to the case with a smaller disorder
strength, but the signal near the mean arrival time ($\tau=0$) is not.
In Fig.~\ref{fig:finitetemp} we show the result for $\Delta_{\xi}=0.1$
at a high temperature $T/\Delta_{\xi}=2000$.  The signal in
a given realization of disorder (thin solid line) is as noisy as in
the zero-temperature case (Fig.~\ref{fig:distrace}), however the
amplitude of the smoothened signal (dashed line) is down by roughly an
order of magnitude.  At this high temperature the disorder-averaged
result (thick solid line) is essentially zero everywhere except very
close to the mean arrival time.

There are several requirements which must be met to make the
measurement useful.  First, one must be able to separately (or at
least differentially) contact the edges of the bilayer system.  If
each electrode used in the measurement couples identically to both
edges, one cannot hope to probe the dynamics of the neutral mode.
Indeed, the sum of the elements of the matrix density-density
correlation function for the 331 state (\ref{eq:DRFull}) is identical
in form to that of the Pfaffian state (\ref{eq:PfDRS}).  The relative
strength of the signal from the neutral mode, compared to the charged
mode, would be maximized by applying a voltage antisymmetric between
the layers.  Second, the experiment must involve a time-resolved
measurement in order to distinguish signals which differ by their
propagation velocities.  Third, the electrodes must be close enough
together so that the decay of the neutral mode signal with distance
does not cause it to be undetectable, but they must be far enough
apart so that the charged and neutral mode signals are well separated
in time.

We believe that the length and time scales needed in a realistic
measurement would require a careful choice of fabrication techniques.
To give more specific estimates, let us take the drift velocity of the
charge mode to be $v_{\rm c}\sim 10^7\,{\rm
cm/s}$\cite{velocity-cleaved,Maasilta-97}.  In a clean system at a
finite temperature the strength of the neutral mode signal decays
exponentially with distance, see Eq.~(\ref{eq:DR331_fint}).  The
separation between the electrodes here cannot be taken to be much
larger than the temperature coherence length $L_T\sim \hbar v/T$,
which is $10^{-2}\,{\rm cm}$ at $10\,{\rm mK}$.  Assuming the neutral
mode velocity is smaller than the charged mode velocity by an order of
magnitude, the arrival time difference between the two modes is on the
order of $10\,{\rm ns}$.  In the presence of disorder, the temperature
coherence length looses its importance, as one can see from the second
term in Eq.~(\ref{eq:DR331_dirt_fint}).  In this case, let us assume a
mean free path of $\ell\sim 1/\Delta_\xi\sim 10^{-5}\,{\rm cm}$, and
require that the electrodes need to be approximately $100\ell\sim
10\,\mu{\rm m}$ apart for the neutral mode signal to be detectable.
The arrival time difference between the signals from the neutral and
charged modes is then around $1\,{\rm ns}$.  If an experimental
measurement like the one described here detected the neutral mode
signal, it would conclusively show that the $\nu=1/2$ plateau in
bilayer systems is the 331 state rather than the Pfaffian.

\section{Details of the Calculations}
\label{sec:details}

In this section we present the detailed calculations of the results
summarized in the previous section.  We begin with a review of the
clean edge theory (Section~\ref{ss:review}), and then discuss the
addition of a parallel magnetic field
(Section~\ref{ss:parallelfield}) and disorder
(Section~\ref{ss:disorder}).  Finally we discuss the Pfaffian edge
(Section~\ref{ss:pfaffian}) and the numerical computations performed
for the 331 edge (Section~\ref{ss:numerics}).

\subsection{Edge Theory of Clean Bilayer Systems}
\label{ss:review}

In this section we review the edge theory of clean bilayer quantum
Hall systems with interlayer electron tunneling.  For a more detailed
discussion see {\bf I}.  The edge theory corresponding to
the Halperin state (\ref{eq:wavefunc}) contains two chiral Bose fields,
$u_i(t,x)\,(i=1,2)$, with compactification radii $R_i=1$
(\ie $u\approx u+2\pi$), and equal-time commutation relations:
\beq
\label{eq:u_CR}
[u_i(t,x),u_j(t,x')]=i\pi K_{ij}\sgn(x-x'),
\eeq
where $K$ is a symmetric, integer-valued matrix which characterizes
the topological properties of the edge and is completely determined by
the exponents in the bulk wavefunction\cite{wen-zee}
\beq
\label{eq:K}
K=\pmatrix{ m & n \cr n & m \cr}.
\eeq
In terms of these fields we can write the charge density and electron
creation operators as
\beq
\label{eq:rho,Psi}
\rho_i(x)={1\over 2\pi}K^{-1}_{ij}\delx u_j(x),\qquad
\Psi_{i}^{\dagger}(x)={1\ov L^{m/2}} e^{-iu_i(x)},
\eeq
and the Hamiltonian as
\widetext
\beq
\label{eq:H0_u}
{\cal H}_0=\ix \left[{{1\ov 4\pi}V_{ij}\,:\!\delx u_i\delx u_j\!:\,
+\lambda_0\left(
{\,:\!\Psi_1(x)\Psi_2^{\dag}(x)\!:\,+\,{\rm h.c.}}\right)}\right],
\eeq
\narrowtext
where
\beq
\label{eq:V}
V=\pmatrix{ v & g \cr g & v \cr },
\eeq
is a symmetric, positive definite ($g^2<v^2$) matrix which includes
the effect of the confining potentials and interactions at the edge,
and $\lambda_0$ is the interlayer electron tunneling amplitude, which
we take to be real without loss of generality.  The normal ordering is
with respect to the oscillator modes of the bosonic fields.

The Hamiltonian and commutation relations can be simplified by the
transformation:
\beq
\label{eq:rotatebosons}
\pmatrix{ u_1 \cr u_2 \cr}={1\ov\sqrt{2}}
\pmatrix{\sqrt{m+n} & -\sqrt{m-n} \cr \sqrt{m+n} & \sqrt{m-n} \cr}
\pmatrix{\phi_{\rm c} \cr \phi_{\rm n} \cr},
\eeq
in terms of which we have
\beq
\label{eq:phi_CR}
[\phi_{\rm i}(x),\phi_{\rm j}(x')]=i\pi\delta_{\rm ij}\sgn(x-x'),
\eeq
and
\widetext
\beq
\label{eq:H0_phi}
{\cal H}_0=\ix\left[{{1\ov 4\pi}v_{\rm c}:(\delx\phi_{\rm c})^2:
+{1\ov 4 \pi}v_{\rm n}:(\delx\phi_{\rm n})^2:+
{2\lambda \ov (2\pi a)^{\bh^2/2}}\cos\left({\bh\phi_{\rm n}}\right)}\right],
\eeq
\narrowtext
where in $\phi_{\rm i}$, the index i runs over the two values i=c,\,n,
which denote the charged and neutral modes, respectively, and we have
introduced the parameters $\bh\equiv\sqrt{2(m-n)}$, $\lambda\equiv
\lambda_0 L^{-n}$, the velocities $v_{\rm c,n}=(m\pm n)(v\pm g)$, and
the short distance cutoff $a$.  The Hamiltonian separates into a free
chiral boson Hamiltonian for the charged mode and a chiral sine-Gordon
Hamiltonian for the neutral mode.  

For future reference we record the expression for the electron and
density operators in terms of the newly introduced bosons
\beqarr
\Psi_{1,2}(x)&=&{1\ov L^{m/2}}e^{i\sqrt{(m+n)/2}\phi_{\rm c}(x)}
e^{\mp i\bh\phi_{\rm n}(x)/2}, 
\label{eq:psi}\\
\rho_{1,2}(x)&=&{1\ov 2\pi\sqrt{2(m+n)}}\delx\phi_{\rm c}(x)
\mp {1\ov 2\pi\bh}\delx\phi_{\rm n}(x)\label{eq:rho},
\eeqarr
which follow from Eqns.~(\ref{eq:rho,Psi}) and (\ref{eq:rotatebosons}).
In the remainder of the paper we will suppress the subscript on the
neutral boson, \ie $\phi\equiv \phi_{\rm n}$.  

The time-ordered correlation functions given in
Section~\ref{ss:reviewS} follow from Eqns.~(\ref{eq:H0_phi}),
(\ref{eq:psi}), and (\ref{eq:rho}); for details see
{\bf I}.  To transform from the time-ordered correlation
functions to the retarded response functions we note that if the
time-ordered function is expressed as 
\beq
\label{eq:CTO}
C(t)=\theta(t)C^>(t)+\theta(-t)C^<(t),
\eeq
where $\theta(t)$ is the Heaviside step function,
then the corresponding retarded correlation function is
\beq
\label{eq:CR}
C^R(t)=\theta(t)[C^>(t)-C^<(t)].
\eeq

\subsection{Parallel Field}
\label{ss:parallelfield}

We consider a parallel magnetic field along the $y$-axis: ${\bf
B}=B\hat{\bf y}$.  This corresponds to a vector potential ${\bf
A}(z)=Bz{\hat{\bf x}}$, where we take the origin of the $z$-axis
midway between the layers, whose separation is $d$.  We incorporate
this parallel field into our edge theory by modifying the charge
density operator via the replacement
\beq
\label{eq:rho/w/A}
\rho_i(x)\mapsto \rho_i(x)-{1\ov 2\pi}A^x(z_i),
\eeq
where $z_{1,2}=\pm d/2$.  Using this along with the definition of the
charge density (\ref{eq:rho,Psi}), and the transformation
(\ref{eq:rotatebosons}) gives a Hamiltonian
\beq
\label{eq:H/w/B}
{\cal H}_B\equiv{\cal H}_0+\ix \left[{{\Gamma\ov 2\pi}\delx\phi +
{\Gamma^2\ov 4\pi v_{\rm n}}}\right],
\eeq
where $\Gamma\equiv v_{\rm n}Bd\bh/2$.  The second term is a constant,
and thus produces only an overall shift in the energy spectrum, and we
will henceforth ignore it.  The interlayer magnetic field
couples only to the neutral mode, and therefore we will not write the
terms involving the charged mode explicitly except when considering
correlation functions.  

As remarked in Section~\ref{ss:parallelfieldS}, the analysis here also
applies to the case where we introduce an electric potential
difference between the layers instead of an interplane magnetic field.
A potential difference $V_{\rm e}$ between the layers adds a term
$V_{\rm e}(\rho_1(x)-\rho_2(x))\approx V_{\rm e}\delx\phi$ to the
Hamiltonian.  This is the same form as the interplane field
perturbation (\ref{eq:H/w/B}).  The only difference is that in the
case of a potential difference between the layers the density
operators are not modified as they are in the interplane magnetic
field case (\ref{eq:rho/w/A}).  However, since ${\cal D}_{ij}(t,x)$
involves the density fluctuation operator,
$\rho_i-\langle\rho_i\rangle$, see Eq.~(\ref{eq:rho_rho}), our results
below for the density-density correlation function apply for either a
parallel magnetic field or an electric potential difference.

\subsubsection{110 Sequence}

The states in the 110 sequence correspond to $\bh^2=2$.  In
{\bf I} it is shown that at this value of $\bh$ the radius
of the neutral boson is $R_{\rm n}=1/\sqrt{2}$ and therefore we can
define a triplet of $\widehat{su}(2)_1$ Kac-Moody (KM) currents
\beqarr
\label{eq:su(2)_curr}
J^z(x)&=&{1\ov 2\pi\sqrt{2}}\delx\phi(x),\nonumber\\
J^{\pm}(x)&=&J^x \pm i J^y={1\ov 2\pi a}e^{\mp i\sqrt{2}\phi(x)},
\eeqarr
in terms of which the Hamiltonian reads
\beq
\label{eq:H/w/B/curr1}
{\cal H}_B=\ix \!\!
\left[{{2\pi v_{\rm n}\ov 3}:\!\left[{{\bf J}(x)}\right]^2\!:\,
+\sqrt{2}\Gamma J^z(x)+2\lambda J^x(x)}\right],
\eeq
where we have used the identity
\beq
\label{eq:Jz_to_J}
\ix :\!\left[{J^z(x)}\right]^2\!:\,|\gamma\ra=\ix {1\ov 3}
:\!\left[{{\bf J}(x)}\right]^2\!:\,|\gamma\ra,
\eeq
valid for any state $|\gamma\ra$ in the Hilbert space.  Next, we can
define a new set of currents ${\tilde J}^a(x)\equiv R^{ab}J^b(x)$,
which also obey an $\widehat{su}(2)_1$ algebra provided $R\in SO(3)$.
In particular if we choose
\beq
\label{eq:R}
R=\pmatrix{\cos\alpha & 0 & -\sin\alpha \cr
0 & 1 & 0 \cr \sin\alpha & 0 & \cos\alpha \cr},
\eeq
where $\sin\alpha \equiv\lambda/\sqrt{\lambda^2+\Gamma^2/2}$,
and express the rotated $\widehat{su}(2)_1$ currents in terms of a
new radius $R=1/\sqrt{2}$ chiral boson, $\theta(x)$, we then have
\widetext
\beqarr
\label{eq:H/w/B/curr2}
{\cal H}_B&=&\ix \!\!\!
\left[{2\pi v_{\rm n}:\!\left[{{\tilde J}^z(x)}\right]^2\!:\,
+2\sqrt{\lambda^2+\Gamma^2/2}\,{\tilde J}^z(x)}\right]\\
&=&\ix\!\!\!\left[{{1\ov 4\pi}v_{\rm n}:\!(\delx \theta)^2\!:\,+
{1\ov \pi\sqrt{2}}\sqrt{\lambda^2+\Gamma^2/2}\,\delx\theta}\right].
\eeqarr
\narrowtext\noindent
Note that this final form of the Hamiltonian is identical to the
neutral mode Hamiltonian in the absence of a parallel magnetic field
with the replacement 
\beq
\label{eq:lambdaprime}
\lambda\mapsto \lp\equiv\sqrt{\lambda^2 +\Gamma^2/2}.
\eeq

The above diagonalization of the Hamiltonian also allows us to find
correlation functions.  In particular, consider the two-point function
of the density-fluctuation operator, ${\cal D}_{ij}(t,x)$ (\ref{eq:rho_rho}).
Using the minimal coupling prescription (\ref{eq:rho/w/A}), the
transformation (\ref{eq:R}), the definitions of the charge density
(\ref{eq:rho}), and $\widehat{su}(2)_1$ currents
(\ref{eq:su(2)_curr}), we can write
\widetext
\beq
\label{eq:rho_in_J}
\rho_{1,2}(x)={1\ov\sqrt{8\pi^2(2m-1)}}\delx \phi_{\rm c}
\mp{1\ov 2\pi}\left[{{\Gamma\ov v_{\rm n}\sqrt{2}}+
{1\ov\lp}\left\{{{\Gamma\ov 2}\delx\theta
-{\lambda\ov a}\cos(\sqrt{2}\theta)}\right\}}\right].
\eeq

Using this expression for the charge density operators in terms of the
fields $\phi_{\rm c}$, and $\theta$, along with the Hamiltonian 
(\ref{eq:H/w/B/curr2}), we can readily find
\beqarr
-i{\cal D}(t,x)&=&{1\ov 2(2m-1)}{(\one+\sigma^x)\ov [\X {\rm
    c}]^2}\nonumber\\ 
&&{}+{1\ov 2 (\lp)^2}{(\one-\sigma^x)\ov [\X {\rm n}]^2}
\left\{{{\Gamma^2\ov 2}+\lambda^2\cos\left({{2\lp
\ov v_{\rm n}}\,x}\right)}\right\}.\label{eq:rho_rho_w/B}
\eeqarr

The evaluation of the single-electron Green's function is more
involved because the electron operators ($\Psi_i(x)$) cannot be
expressed in terms of the fields $\phi_{\rm c}$ and $\theta$.
Following the method used in {\bf I}, we can use the
independence of the charged and neutral modes to write the Green's
function, ${\cal G}_{ij}(t,x)$ (\ref{eq:prop}), for all states in the
110 sequence as
\beq
\label{eq:ratiotrick}
{\cal G}_{ij}(t,x)={\cal G}^{(110)}_{ij}(t,x){{\cal G}_{\phi_{\rm
c}}^{(m)}(t,x)\ov {\cal G}_{\phi_{\rm c}}^{(1)}(t,x)},
\eeq
where ${\cal G}^{(110)}_{ij}$ is the Green's function for the special
case $m=1,\,n=0$ and 
\beq
\label{eq:Gcharged}
{\cal G}_{\phi_{\rm c}}^{(m)}(t,x)\equiv{1\ov L^{m-1/2}}
\left\la{e^{i\sqrt{m-{1\over 2}}\phi_{\rm c}(t,x)}e^{-i\sqrt{m-{1\over
2}}\phi_{\rm c}(0,0)}}\right\ra={1\ov [\X {\rm c}]^{m-{1\ov 2}}}.
\eeq
The decomposition (\ref{eq:ratiotrick}) is useful because for the
uncorrelated integer 110 state there exists a chiral fermion
description of the edge theory including tunneling and a parallel
field:
\beq
\label{eq:HB110}
{\cal H}^{(110)}_B=\ix:\!\left[{-iv\psi_i^{\dagger}\delx\psi_i +2\pi g
\psi_1^{\dagger}\psi_1 \psi_2^{\dagger}\psi_2 
-\lambda(\psi_2^{\dagger}\psi_1
+\psi_1^{\dag}\psi_2)-{\Gamma\ov\sqrt{2}}
(\psi_1^{\dag}\psi_1-\psi_2^{\dag}\psi_2)
}\right]\!:.
\eeq
If we perform the following canonical transformation
\beq
\label{eq:sigmay}
\pmatrix{\psi_1 \cr\psi_2}=e^{i\varphi\sigma_y}\pmatrix{\psi_{+}\cr\psi_{-}}
\quad{\rm where}\quad \sin(2\varphi)=-{\lambda\ov\lp},
\eeq
then the Hamiltonian (\ref{eq:HB110}) becomes
\beq
\label{eq:HB1102}
{\cal H}^{(110)}_B=\ix:\!\left[{-iv\left({\psi_+^{\dagger}\delx\psi_+ +
\psi_-^{\dagger}\delx\psi_-}\right) +2\pi g\psi_+^{\dagger}\psi_+ 
\psi_-^{\dagger}\psi_- -\lp(\psi_+^{\dagger}\psi_+ -\psi_-^{\dagger}\psi_-)
}\right]\!:.
\eeq
By bosonizing according to $\psi_{\pm}(x)=e^{i\phi_{\pm}(x)}/\sqrt{2\pi a}$,  
and defining $\theta_+\equiv(\phi_+ +\phi_-)/\sqrt{2}$, and 
$\theta_-\equiv(\phi_+-\phi_-)/ \sqrt{2}-{\sqrt{2}\lp x/ v_{\rm n}}$,
we can exactly evaluate the single-electron Green's function
\beq
\label{eq:exactG110}
{\cal G}^{(110)}(t,x)={1\ov 2\pi\sqrt{(x-v_{\rm c}t+i\epsilon_t)
(x-v_{\rm n}t+i\epsilon_t)}}
\exp\left[{i\left({{\Gamma\ov\sqrt{2}}\sigma^z
+\lambda\sigma^x}\right){x\ov v_{\rm n}}}\right].
\eeq
Combining Eqns.~(\ref{eq:ratiotrick}), (\ref{eq:Gcharged}), and
(\ref{eq:exactG110}), we finally arrive at
\beqarr
{\cal G}(t,x)&=&{1\ov [\X {\rm c}]^{m-{1\ov 2}}\sqrt{\X {\rm
n}}}\nonumber\\
&&{}\times\left[{\one\cos(\lp x/v_{\rm n}) +{i\ov\lp}
\left({{\Gamma\ov\sqrt{2}}\sigma^z+\lambda\sigma^x}\right)
\sin(\lp x/v_{\rm n})}\right]\label{eq:G_B}.
\eeqarr

\subsubsection{331 Sequence}  

The states in the 331 sequence correspond to $\bh^2=4$.  In
{\bf I} it is shown that at this value of $\bh$ we can
fermionize the neutral boson using
\beq
\label{eq:fermi_331}
{1\ov \sqrt{2\pi a}}e^{i\phi}=\psi,\quad
{1\ov 2\pi}\delx\phi=\,:\!\psi^{\dag}\psi\!:,\quad
{i\ov 2\pi a^2}e^{i2\phi}=\psi\delx\psi.
\eeq
With these identities, the neutral mode part of the Hamiltonian 
(\ref{eq:H/w/B}) becomes
\beqarr
{\cal H}_B&=&\ix:\!\left[{-iv_{\rm n}\psi^{\dag}\delx\psi
-i{\lambda\ov 2\pi}(\psi^{\dag}\delx\psi^{\dag}+\psi\delx\psi)
+\Gamma\psi^{\dag}\psi}\right]\!:\label{eq:H/w/B/331/D}\\
&=&\ix:\!\left[{-{i\ov 2}v_1\chi_1\delx\chi_1-{i\ov 2}v_2\chi_2\delx\chi_2
+i\Gamma\chi_1\chi_2}\right]\!:,\label{eq:H/w/B/331/M}
\eeqarr
\narrowtext
where we have written the chiral Dirac fermion in terms of its
Majorana components: $\psi(x)=[\chi_1(x)+i\chi_2(x)]/\sqrt{2}$, where
$\chi_i^{\dag}=\chi_i$, and recalled $v_{1,2}= v_{\rm n}\pm
\lambda/\pi$.  The tunneling term splits the velocities of the two
Majorana fermions and the parallel field term couples them.

The Hamiltonian is quadratic and hence readily diagonalizable.  If we
take antiperiodic boundary conditions for the Fermi field,
$\psi(x+L)=-\psi(x)$, and expand in Fourier modes according to
\beq
\label{eq:psi_modes}
\psi(x)={1\ov\sqrt{L}}\sum_k e^{ikx}c_k,
\eeq
where $k\in (2\pi/L)(\Zint+1/2)$, the Hamiltonian (\ref{eq:H/w/B/331/D}) 
becomes
\beqarr
{\cal H}_B&=&\sum_k:\!\left[{v_{\rm n}kc_k^{\dag}c_k+\Gamma
c_k^{\dag}c_k + {\lambda k\ov 2\pi}(c_k^{\dag}c_{-k}^{\dag}
-c_k c_{-k})}\right]\!:\nonumber\\
& &\hskip-0.4in=\sum_{k>0}:\!\pmatrix{c_k^{\dag} & c_{-k} \cr}
\pmatrix{v_{\rm n}k+\Gamma & \lambda k/\pi \cr
\lambda k/\pi & v_{\rm n}k-\Gamma \cr}
\pmatrix{c_k \cr c_{-k}^{\dag}}\!:.\label{eq:H/w/B/modes}
\eeqarr
Employing the Bogoliubov transformation 
\beq
\label{eq:bogo}
\pmatrix{c_k \cr c_{-k}^{\dag}}=
\pmatrix{\cos\alpha_k & -\sin\alpha_k \cr
\sin\alpha_k & \cos\alpha_k \cr}\pmatrix{a_k \cr a_{-k}^{\dag} \cr},
\eeq
where $\alpha_{-k}=-\alpha_k$, the Hamiltonian is diagonalized via the
choice
\beq
\label{eq:alpha}
\tan(2\alpha_k)={\lambda k\ov \pi\Gamma},
\eeq
along with the restriction $2\alpha_k \in [-\pi/2,\pi/2]$, required to
produce the correct spectrum in the limit $\lambda\rightarrow 0$. 
This yields
\beqarr
\label{eq:H/w/B/diag}
{\cal H}_B&=&\sum_k\left({v_{\rm n}k + 
\sgn(\Gamma)\sqrt{(\lambda
  k/\pi)^2+\Gamma^2}}\right)\,:\!a_k^{\dag}a_k\!:\nonumber\\ 
&\equiv&\sum_k\eps(k)\,:\!a_k^{\dag}a_k\!:.
\eeqarr

To calculate correlation functions, we first use the transformation
(\ref{eq:bogo}), along with the expressions for $\alpha_k$
(\ref{eq:alpha}) and $\psi(x)$ (\ref{eq:psi_modes}) to express the
Fermi field in terms of the mode operators which diagonalize the
Hamiltonian
\widetext
\beqarr
\psi(x)&=&{1\ov\sqrt{2L}}\sum_k{e^{ikx}\ov[(\lambda
k/\pi)^2+\Gamma^2]^{1/4}}\nonumber\\
&&{}\times \left[{\sqrt{\sqrt{(\lambda
k/\pi)^2+\Gamma^2}+|\Gamma|}\,a_k\!-\!
\sgn(\Gamma\lambda k)\sqrt{\sqrt{(\lambda
k/\pi)^2+\Gamma^2}-|\Gamma|}\,a_{-k}^{\dag}}\right]\!.\label{eq:psi_a}
\eeqarr
\narrowtext
To compute the single-electron Green's function (\ref{eq:prop})
we use the transformation (\ref{eq:rotatebosons}) and the fermionization
(\ref{eq:fermi_331}) to write the electron operators (\ref{eq:rho,Psi}) as
\beq
\label{eq:elec_331}
:\!\Psi_i\!:\, ={1\over (2\pi a)^{(m-1)/2}} e^{i\sqrt{m-1}\phi_{\rm c}}
\left({\delta_{i1}\psi^{\dagger}+\delta_{i2}\psi}\right).
\eeq

Since the charged and neutral modes are not coupled, we find
\beq
\label{eq:prop_331_0}
{\cal G}(t,x)={1\over \left[{\X {\rm c}}\right]^{m-1}}{\cal
G}_{\psi}(t,x),
\eeq
where we have defined the matrix
\beq
\label{eq:prop_331_1}
{\cal G}_{\psi}(t,x)\equiv -i
\pmatrix{\la T\psi(t,x)\psi^{\dag}(0,0)\ra & \la
T\psi(t,x)\psi(0,0)\ra \cr \la T\psi^{\dag}(t,x)\psi^{\dag}(0,0)\ra &
\la T\psi^{\dag}(t,x)\psi(0,0)\ra \cr}.
\eeq

The correlation function of the fermionized neutral mode can be
reduced to quadrature.  Using Eqns.~(\ref{eq:H/w/B/diag}) and 
(\ref{eq:psi_a}) we obtain (in the limit $L\rightarrow \8$)
\widetext
\beq
\label{eq:G_331_B}
{\cal G}_{\psi}(t,x)={-i\ov 4\pi}
\pmatrix{ 1+(i/t)(\del/\del\Gamma) & (\lambda/\pi\Gamma t)(\del^2/\del
x\del \Gamma) \cr (\lambda/\pi\Gamma t)(\del^2/\del
x\del \Gamma) &  1-(i/t)(\del/\del\Gamma)
\cr}I,
\eeq
where there remains the integral
\beq
\label{eq:integral}
I\equiv \int dk\,\left[{
\sgn(t)\cos(kx-\eps(k)t)+i\sgn(k-k_F)\sin(kx-\eps(k)t)}\right],
\eeq
and the ``Fermi momentum'', defined by $\eps(k_F)=0$, is given by 
$k_F=-\Gamma/\sqrt{v_1v_2}$.  To simplify the result, we first change 
the variable of integration to $\omega\equiv \eps(k)$ for $k>k_F$ and 
$\omega\equiv -\eps(k)$ for $k<k_F$.  With these substitutions and
some algebra, the matrix Green's function can be written
\beqarr
{\cal G}_{\psi}(t,x)&=&{-i\ov v_1v_2}\sgn(t)\biggl[{
\left({v_{\rm n}\one -{\lambda\ov\pi}\sigma^x}\right)P_X(\tau,X)
+ \sgn(t)\left({{\lambda\ov\pi}\one-v_{\rm n}\sigma^x}\right){\lambda\ov\pi}
P_{\tau}(\tau,X)}\nonumber\\
&&{}-iv_1v_2\Gamma \sigma^zP(\tau,X)\biggr]\label{eq:G_psi_B}
\eeqarr
\narrowtext
where we have defined $X\equiv x/v_1v_2$, $\tau\equiv \sgn(t)(v_{\rm
n}X-t)$, and the function
\beq
\label{eq:P}
P(\tau,X)\equiv\int_0^{\8}{d\omega\ov 2\pi}e^{i\omega\tau}
{\sin(\kappa(\omega)X)\ov \kappa(\omega)},
\eeq
where $\kappa(\omega)\equiv\sqrt{v_1v_2\Gamma^2
+(\lambda/\pi)^2\omega^2}$, and the subscripts on $P$ in
Eq.~(\ref{eq:G_psi_B}) denote partial differentiation.  Although we
have been unable to evaluate $P(\tau,X)$ explicitly, the real part of
this function can be calculated in closed form.  This is discussed
below in Section~\ref{ss:disorder} when we consider the retarded
version of ${\cal G}_{\psi}$.

Turning now to the density-density correlation function, we can use
the fermionization (\ref{eq:fermi_331}) to write the density operators
(\ref{eq:rho}) as
\beq
\label{eq:rho331}
\rho_{1,2}={1\ov 4\pi\sqrt{m-1}}
\delx\phi_{\rm c}\mp{1\ov 2}\,:\!\psi^{\dag}\psi\!:.
\eeq
Since the Hamiltonian in terms of $\psi$ (\ref{eq:H/w/B/331/D}) is quadratic,
Wick's theorem holds for this field and the density two-point function
can be expressed in terms of the single-particle Green's function.
Using (\ref{eq:rho331}) and (\ref{eq:rho_rho}) one finds
\widetext
\beq
\label{eq:DDw/B,331}
-i{\cal D}(t,x)={1\ov 4(m-1)}{(\one+\sigma^x)\ov [\X {\rm c}]^2}
+{(\one-\sigma^x)\ov 4}\det\,{\cal G}_{\psi}(t,x).
\eeq
Along with Eq.~(\ref{eq:G_psi_B}), this gives an expression for the
density-density correlation function in terms of the single function
$P(\tau,X)$ and its derivatives:
\beq
\label{eq:DDw/B,331,2}
-i{\cal D}(t,x)={1\ov 4(m-1)}{(\one+\sigma^x)\ov [\X {\rm c}]^2}
-{(\one-\sigma^x)\ov 4v_1v_2}\left[{P_X^2-{\lambda^2\ov \pi^2}P_{\tau}^2
+v_1v_2\Gamma^2P^2}\right].
\eeq
\narrowtext

\subsection{Disorder}
\label{ss:disorder}

We now consider the effects of adding several types of disorder to the
Hamiltonian of the bilayer system (\ref{eq:H0_u}).  We begin by
considering the relevancy of various random terms within a
renormalization group (RG) analysis.  We then present exact results
for the 110 and 331 sequences, concentrating on the retarded density
response function because of its relevancy for experiments.  The 110
case is solved by using an $SU(2)$ gauge transformation to separate
the Green's functions into products of clean Green's functions and
terms involving only the random fields.  After this step, the
disordered problem is shown to be equivalent to a spin-1/2 particle in
a random magnetic field and the disorder averaging is performed
non-perturbatively.  The solution for the 331 sequence involves an
exact summation of the disorder-averaged perturbation theory which is
possible because of the chirality of the system.  In
Appendix~\ref{sec:331-via-spin} we present an alternate method for
obtaining disorder-averaged correlation functions for the 331 sequence
based on the spin analogy.

If we consider a general perturbation to the Hamiltonian of the form
\beq
\label{eq:general_dirt}
{\cal H}_{\delta}=\ix \zeta(x){\cal O}(x),
\eeq
where ${\cal O}(x)$ is an operator of scaling dimension $\delta$ and
$\zeta(x)$ is a Gaussian random variable with variance $\Delta$, \ie 
$\ol{\zeta(x)\zeta(x')}=\Delta\delta(x-x')$, where the bar denotes
disorder averaging, then a lowest-order perturbative RG analysis
gives\cite{g-s}
\beq
\label{eq:RG}
{d\Delta\ov d\ell}=(3-2\delta)\Delta,
\eeq
where the short distance cutoff increases as $\ell$ increases.

Consider first the possibility of disorder in the velocity-interaction
matrix $V$ in the Hamiltonian (\ref{eq:H0_u}).  Since the $V$ matrix
multiplies an operator of scaling dimension $\delta=2$, we see from
the flow equation (\ref{eq:RG}) that delta-correlated disorder is
RG irrelevant.  Therefore we can ignore randomness in the $V$ matrix
and interpret the values appearing in Eq.~(\ref{eq:V}) as
disorder-averaged mean values.  Note that the $V$ matrix must be
symmetric, since it multiplies a symmetric operator in the Hamiltonian,
and the positive-definiteness of $V$ is required for the
Hamiltonian to be bounded from below.  However, the assumption that
$V_{11}=V_{22}$ is made for technical reasons and it can be relaxed to
$\ol{V}_{11}=\ol{V}_{22}$, a weaker criterion.

Next we turn to the case of disorder in the tunneling amplitude
$\lambda$ in the Hamiltonian (\ref{eq:H0_phi}).  The scaling dimension
of the tunneling operator it multiplies is $\delta=\bh^2/2$.  Using
Eq.~(\ref{eq:RG}), we find that disorder in the tunneling amplitude is
relevant for the 110 sequence (\ie $\bh^2=2$), and irrelevant for
the 331 sequence (\ie $\bh^2=4$).  The former case will be discussed
later in this section and the latter case in 
Appendix~\ref{sec:random-tunneling}.

Finally we consider adding random scalar potentials which couple to
the edge charge densities in each layer:
\widetext
\beq
\label{eq:scalar_dirt1}
{\cal H}_{\xi}=\ix \left[{\xi_1(x)\rho_1(x)+\xi_2(x)\rho_2(x)}\right]
=\ix \left[{\xi_{\rm c}(x){1\ov 2\pi}\delx\phi_{\rm c}(x)+
\xi_{\rm n}(x){1 \ov 2\pi}\delx\phi(x)}\right],
\eeq
where we have used the definition of the charge density operators
(\ref{eq:rho}), and
\beq
\label{eq:scalar_dirt2}
\xi_{\rm c}\equiv {1\ov \sqrt{2(m+n)}}(\xi_1+\xi_2),\qquad
\xi_{\rm n}\equiv {1\ov \bh}(\xi_2-\xi_1).
\eeq
If we assume that $\xi_{1,2}(x)$ are independent Gaussian random
variables then so are $\xi_{\rm c,n}(x)$.  Since these terms involve
operators of scaling dimension $\delta=1$ they are relevant
perturbations for both the 110 and 331 sequences.

\subsubsection{110 Sequence}

In the discussion above we found that for the 110 sequence
($\bh^2=2$), disorder both in the tunneling and in the scalar
potential terms is relevant.  We therefore consider the Hamiltonian
\beqarr
{\cal H}_D&=&\ix \left[{{v_{\rm c}\ov 4\pi}\,:\!(\delx \phi_{\rm c})^2\!:\,
+{v_{\rm n}\ov 4\pi}\,:\!(\delx \phi)^2\!:\,+ {1\ov 2\pi a}
\left({\lambda(x)e^{i\sqrt{2}\phi(x)}+\lambda^{*}(x)
e^{-i\sqrt{2}\phi(x)}}\right)}\right.\nonumber\\
&&\left.{{}+\xi_{\rm c}(x){1\ov 2\pi}\delx\phi_{\rm
c}(x)+ \xi_{\rm n}(x){1 \ov
2\pi}\delx\phi(x)}\right]\label{eq:H110_w/dirt},
\eeqarr
\narrowtext
where $\lambda(x)$ is a complex random tunneling amplitude.  The
presence of disorder breaks translation invariance and hence the
current algebra method used to solve the clean problem in
Section~\ref{ss:parallelfield} is of no use because the transformation
${\tilde J}^a(x)=R^{ab}J^b(x)$ must be a global rotation to map
between sets of KM generators.  However, an alternate approach to the
problem developed in {\bf I} is useful in
the disordered case.  We add to the Hamiltonian (\ref{eq:H110_w/dirt})
an auxiliary free chiral boson (${\ph}$) with a velocity equal to the
velocity of $\phi$:
\beq
\label{eq:addauxboson}
{\cal H}_D\mapsto {\cal H}_D+\ix {1\ov 4\pi}v_{\rm n}\,:\!(\delx\ph)^2\!:,
\eeq
perform the canonical transformation
\beq
\label{eq:rotatebosons2}
\pmatrix{\ph \cr \phi \cr}={1\over \sqrt{2}}\pmatrix{1 & 1
\cr 1 & -1 \cr} \pmatrix{\theta_1 \cr \theta_2},
\eeq
and then fermionize according to
\beq
\label{eq:fermi_theta}
\psi_i(x)={1\over \sqrt{2\pi a}}e^{i\theta_i (x)}.
\eeq
The details required to make this mapping rigorous
(\ie compactification radii, topological charges, Klein factors) are
discussed in {\bf I}.  The result of this procedure is a
quadratic Hamiltonian
\widetext
\beq
\label{eq:H110_w/dirt_quad}
{\cal H}_D=\ix \left[{{1\ov 4\pi}v_{\rm c}\,:\!(\delx \phi_{\rm c})^2\!:\,
+\xi_{\rm c}(x){1\ov 2\pi}\delx\phi_{\rm c}(x)
+\,:\!\left({-iv_{\rm n}\Psi^{\dag}\delx{\Psi}+v_{\rm
      n}B^a(x)\Psi^{\dag}\sigma^a 
\Psi}\right)\!:}\right],
\eeq
where we have defined
\beq
\label{eq:Psi,B}
\Psi(x)\equiv\pmatrix{\psi_1(x)\cr\psi_2(x)},\qquad 
{\bf B}(x)={1\ov v_{\rm n}}(-\Re e\, [\lambda(x)],- \Im m[\lambda(x)],
\xi_{\rm n}(x)/\sqrt{2}),
\eeq
\narrowtext
and the index $a$ runs over $x,y,z$.  The fermionic part of this
Hamiltonian describes a pseudo-spin-1/2 fermion coupled to a random
$SU(2)$ gauge field.

We now perform a change of variables that absorbs the disordered
terms into the definitions of the field operators.  For the charged
mode we define
\beq
\label{eq:eta_c}
\eta(x)=\phi_{\rm c}(x)+{1\ov v_{\rm c}}\int^x dy\,\xi_{\rm c}(y),
\eeq
and for the neutral mode we use an $SU(2)$ gauge transformation
\beq
\label{eq:SU(2)rot}
\Psi(x)=S(x)\Pt(x),
\eeq
where $S(x)\in SU(2)$ is a solution of the matrix differential
equation
\beq
\label{eq:diffeq}
{dS(x)\ov dx}=-iB^a(x)\sigma^a S(x).
\eeq
With these definitions, the Hamiltonian (\ref{eq:H110_w/dirt_quad}) becomes
\beq
\label{eq:H110_quad}
{\cal H}_D=\ix \left[{{1\ov 4\pi}v_{\rm c}\,:\!(\delx\eta)^2\!:\,
-{1\ov 4\pi v_{\rm c}}\xi_{\rm c}^2
-iv_{\rm n}\,:\!\Pt^{\dag}\delx\Pt\!:\,}\right].
\eeq
Since the second term only involves the disordered scalar potential,
it does not affect correlation functions and will henceforth be
neglected.  Note that in going from Hamiltonian
(\ref{eq:H110_w/dirt_quad}) to Hamiltonian (\ref{eq:H110_quad}) we
have used a gauge transformation on a chiral Fermi field
(\ref{eq:SU(2)rot}) without accounting for the chiral anomaly.  This
is valid because the gauge field is a quenched random variable; in
this case the anomaly associated with the chiral gauge transformation
(\ref{eq:SU(2)rot}) cancels in the average.

Our primary goal is to understand the behavior of the density-density
correlation function in a given sample, \ie for a given realization
of disorder.  We begin by expressing the density operators
(\ref{eq:rho}) in terms of the fields $\eta$ and $\Psi$, with the help
of Eqns.~(\ref{eq:rotatebosons2}), (\ref{eq:fermi_theta}), and
(\ref{eq:eta_c}):
\widetext
\beq
\label{eq:rho_110}
\rho_{1,2}(x)={1\ov 2\pi\sqrt{2(2m-1)}}\left({\delx\eta(x)-{1\ov
v_{\rm c}}\xi_{\rm c}(x)}\right)\mp {1\ov
2}\,:\!\Psi^{\dag}(x)\sigma^z \Psi(x)\!:.
\eeq
Using this expression in the definition of the density two-point
function (\ref{eq:rho_rho}) we have
\beq
\label{eq:rho_rho_110}
-i{\cal D}(t,x,x')\!=\!{1\ov 2(2m-1)}{(\one+\sigma^x)\ov [\XP {\rm c}]^2}
-{(\one-\sigma^x)\ov 4}\tr\left({\sigma^z G(t,x,x')\sigma^z G(-t,x',x)}\right)
\eeq
\narrowtext
where we have used the single-particle matrix Green's function
\beq
\label{eq:G110}
G_{ij}(t,x,x')=-i\la T\psi_i(t,x)\psi_j^{\dag}(0,x')\ra.
\eeq
We have explicitly included two spatial arguments in these correlation
functions because of the lack of translation invariance in a given
realization of disorder.  Note that although the charged mode,
$\phi_{\rm c}$, is coupled in (\ref{eq:scalar_dirt1}) to a disorder
potential, $\xi_{\rm c}$, the charged-mode part of the above
correlation function (\ref{eq:rho_rho_110}) is identical to the result
in the absence of disorder.  This result is true for every realization
of disorder and is essentially equivalent to the loop-cancelation
theorem which states that for linearly dispersing fermions
($\eps(k)\propto k$) in 1+1 dimensions the connected $n$-point
function of the density operator vanishes identically for
$n>2$\cite{metzner}.

To determine the effect of disorder on the neutral mode we first note
that the differential equation (\ref{eq:diffeq}) has a solution in
terms of a coordinate-ordered exponential
\beq
\label{eq:S}
S(x)=T_y\exp\left({-i\int^x dy\,B^a(y)\sigma^a}\right),
\eeq
where $T_y$ is the $y$-ordering operator.  Since the matrix $S(x)$ can
be taken outside quantum expectation values, we can express the
Green's function of the $\Psi$ field (\ref{eq:G110}) in terms of the
Green's function of the (free) $\Pt$ field and thus write
\beq
G_{ij}(t,x,x')={1\ov \XP {\rm n}}U_{ij}(x,x'),
\label{eq:GtoGfree}
\eeq
where
\beq
U(x,x')\equiv S(x)S^{\dag}(x')=T_y\exp\left
({-i\int^x_{x'} dy\,B^a(y)\sigma^a}\right)\label{eq:U}
\eeq
is a unitary matrix.  Using Eq.~(\ref{eq:GtoGfree}) we find that the
neutral mode part of the density-density correlation function is
proportional to
\widetext
\beq
\label{eq:D_neutral}
{-\tr\left({\sigma^z G(t,x,x')\sigma^z G(-t,x',x)}\right)}
{1\ov [\XP {\rm n}]^2}\tr\left({\sigma^zU(x,x')\sigma^zU^{\dag}(x,x')}\right),
\eeq
\narrowtext
where we have used the property $U^{\dag}(x,x')=U(x',x)$, which
follows from the definition (\ref{eq:U}).  In Eq.~(\ref{eq:D_neutral})
we have written a correlation function in the disordered system as the
product of the corresponding function in the clean system and a factor
which depends only on the random potential.

If we interpret the coordinate $y$ appearing in the definition of $U$
as a fictitious time, then the matrix $U(x,x')$ is exactly the time
evolution operator between times $x$ and $x'$ for a zero-dimensional
system with time-dependent Hamiltonian $B^a(y)\sigma^a$.  This is the
Hamiltonian for a spin-1/2 object in a random magnetic field $B^a(y)$.
The quantity appearing in the trace in Eq.~(\ref{eq:D_neutral}) can
then be interpreted as the $\la S_z(x)S_z(x')\ra$ correlation function
for this spin.

To understand the behavior of the density-density correlation function
in a given sample we will first calculate its disorder-average, which
involves averaging the quantity appearing on the r.h.s. of
Eq.~(\ref{eq:D_neutral}).  Toward this end, consider the following
vector quantity
\beq
\label{eq:F}
F^a(x;x')=\tr\left({U^{\dag}(x,x')\sigma^aU(x,x')\sigma^z}\right).
\eeq
By differentiating with respect to $x$ we find that this is a solution
of the differential equation
\beq
\label{eq:diffeqF}
{dF^a(x;x')\ov dx}=M^{ab}(x)F^b(x;x'),
\eeq
{where}
$M^{ab}(x)\equiv -2\epsilon^{abc}B^c(x)$,
subject to the boundary condition $F^a(x';x')=\tr(\sigma^a\sigma^z)
=2\delta^{az}$.  The solution of this differential equation can also
be written as
\beq
\label{eq:FfromM}
F^a(x;x')=
\left[{T_y\exp\left({\int^x_{x'}dy\,M(y)}\right)}\right]^{ab}F^b(x';x').
\eeq
We have expressed the quantity we desire, $F^z(x;x')$, in terms of a
single coordinate-ordered matrix exponential, whose disorder average
we now show can be readily evaluated.

We assume the tunneling amplitude and scalar potential are
delta-correlated Gaussian random variables and denote the mean and
variance of $B^a(x)$ by $\mu^a$ and $\Delta^a$, respectively.  Since
the exponential appearing in Eq.~(\ref{eq:FfromM}) is ordered in $y$,
and the elements of $M(y)$ are independently distributed for each $y$,
we can consider breaking up the interval $[x',x]$ into $N$ intervals
of length $\epsilon=|x-x'|/N$ and then taking the limit
$N\rightarrow\8$.  Therefore we can write
\widetext
\beq
\label{eq:disavgM}
\ol{T_y\exp\left({\int^x_{x'}dy\,M(y)}\right)}=
\lim_{N\rightarrow\8}\left[{\int d{\bf B}\,P({\bf B})
e^{\sgn(x-x')\epsilon M}}\right]^N,
\eeq
where the probability distribution is
\beq
\label{eq:probB}
P({\bf B})d{\bf B}=\sqrt{\epsilon^3\ov(2\pi)^3\Delta^x
\Delta^y\Delta^z}\exp\left[{-{1\ov 2}\sum_{a=1}^3
{\epsilon\ov \Delta^a}(B^a-\mu^a)^2}\right]d{\bf B}.
\eeq
Expanding the $e^{\sgn(x-x')\epsilon M}$ factor in Eq.~(\ref{eq:disavgM}) and
performing the integration gives
\beq
\label{eq:disavgM2}
\ol{T_y\exp\left({\int^x_{x'}dy\,M(y)}\right)}=
\lim_{N\rightarrow\8}\left[{1-{2|x-x'|\ov N}W+{\cal O}\left({1\ov
N^2}\right)}\right]^N
=\exp\left({-2|x-x'|W}\right),
\eeq
where 
\beq
\label{eq:W}
W\equiv\pmatrix{ \Delta^y+\Delta^z & 0 & 0 \cr
0 & \Delta^z+\Delta^x & 0 \cr
0 & 0 & \Delta^x+\Delta^y \cr}
+\sgn(x-x')\pmatrix{0 & \mu^z & -\mu^y \cr
-\mu^z & 0 & \mu^x \cr
\mu^y & -\mu^x & 0 \cr}.
\eeq
{}From Eqns.~(\ref{eq:F}), (\ref{eq:FfromM}), and (\ref{eq:disavgM2}),
we finally 
arrive at 
\beq
\label{eq:DisAvgSzSz}
\ol{\tr\left({U^{\dag}(x,x')\sigma^zU(x,x')\sigma^z}\right)}=
2\left[{e^{-2|x-x'|W}}\right]^{zz}.
\eeq

While it is in principle possible to evaluate the exponential of $W$
for arbitrary $\mu^a$ and $\Delta^a$, for simplicity 
we shall restrict ourselves to the case of a real
tunneling amplitude.  If the tunneling amplitude has mean $\lambda$
and variance $\Delta_{\lambda}$, and the disordered scalar potential
has mean zero and variance $\Delta_{\xi}$, then from the definition of
${\bf B}(x)$ (\ref{eq:Psi,B}) we have
\beqarr
\mu^x&=&-{\lambda\ov v_{\rm n}},\quad\mu^y=0,\quad\mu^z=0\nonumber\\
\Delta^x&=&{\Delta_{\lambda}\ov v_{\rm n}^2},\quad \Delta^y=0,\quad 
\Delta^z={\Delta_{\xi}\ov 2v_{\rm n}^2}.\label{eq:mu's_Delta's}
\eeqarr
In this case 
\beq
\label{eq:Wspec}
W=\left[{{\Delta_{\xi}\ov 2v_{\rm n}^2}}\right]\oplus
\left[{\one\,{\Delta_{\lambda}+\Delta_{\xi}/4\ov 
v_{\rm n}^2} + \sigma^z\,{\Delta_{\xi}\ov 4v_{\rm n}^2}
-i\sigma^y\,\sgn(x-x'){\lambda\ov v_{\rm n}} }\right],
\eeq
and from Eqns.~(\ref{eq:rho_rho_110}), (\ref{eq:D_neutral}), and 
(\ref{eq:DisAvgSzSz}) we find
\beqarr
-i\ol{{\cal D}}(t,x,x')&=&{1\ov 2(2m-1)}
{(\one+\sigma^x)\ov [\XP {\rm c}]^2}\nonumber\\
&&{}+{1\ov 2}{(\one-\sigma^x)\ov [\XP {\rm n}]^2}
\exp\left({-{2|x-x'|\ov v_{\rm n}^2}
(\Delta_{\lambda}+\Delta_{\xi}/4)}\right)\nonumber\\
&&{}\times\left[{\cos\left({2|x-x'|\tilde{\lambda}\ov v_{\rm n}}\right)+
{\Delta_{\xi}\ov4v_{\rm n}\tilde{\lambda}}
\sin\left({2|x-x'|\tilde{\lambda}\ov v_{\rm n}}\right)}\right],
\label{eq:DisAvgD}
\eeqarr
where
\beq
\label{eq:shiftlambda}
\tilde{\lambda}\equiv
\sqrt{\lambda^2-\left({\Delta_{\xi}\ov 4v_{\rm n}}\right)^2}.
\eeq
Transforming to the corresponding retarded function using
Eqns.~(\ref{eq:CTO}) and (\ref{eq:CR}) we arrive at
Eq.~(\ref{eq:DR110_fint2S}).

The question remains as to what the behavior of ${\cal D}(t,x,x')$ is
in a given sample.  Instead of attempting to evaluate higher moments
of this correlation function, we shall exploit the fact that it can be
expressed in terms of the single-particle Green's function, whose
second moment we will compute.  Writing out the trace in
Eq.~(\ref{eq:rho_rho_110}) explicitly we find
\beqarr
\!\!\tr\left({\sigma^z G(t,x,x')\sigma^z G(-t,x',x)}\right)
\!=\!G_{11}(t,x,x')G_{11}(-t,x',x)
\!-\!G_{12}(t,x,x')G_{21}(-t,x',x)&&\nonumber\\
-\!G_{21}(t,x,x')G_{12}(-t,x',x)\!+\!G_{22}(t,x,x')G_{22}(-t,x',x).&&
\label{eq:tr_from_G} 
\eeqarr
{}From Eq.~(\ref{eq:GtoGfree}) and the result
\beq
\label{eq:DisAvgU}
\ol{U(x,x')}=\exp\left({-{|x-x'|\ov 2v_{\rm n}^2}
(\Delta_{\lambda}+\Delta_{\xi}/2)}\right)
\left[{\one\cos\left({{\lambda\ov v_{\rm n}}(x-x')}\right)
+i\sigma^x\sin\left({{\lambda\ov v_{\rm n}}(x-x')}\right)
}\right],
\eeq
which can be obtained via the same procedure used to evaluate the
average in Eq.~(\ref{eq:disavgM}), we find the disorder-averaged
single-particle Green's functions are also exponentially decaying in
space
\beqarr
\ol{G}_{11}(t,x,x')=\ol{G}_{22}(t,x,x')&=&
{e^{-|x-x'|(\Delta_{\lambda}+\Delta_{\xi}/2)/2v_{\rm n}^2}
\ov \XP {\rm n}}\cos\left({{\lambda\ov v_{\rm
n}}(x-x')}\right),
\label{eq:DisAvgG11}\\
\ol{G}_{12}(t,x,x')=\ol{G}_{21}(t,x,x')&=&i
{e^{-|x-x'|(\Delta_{\lambda}+\Delta_{\xi}/2)/2v_{\rm n}^2}
\ov \XP {\rm n}}\sin\left({{\lambda\ov v_{\rm
n}}(x-x')}\right).
\label{eq:DisAvgG12}
\eeqarr
To investigate whether this exponential decay is an artifact of the
disorder averaging (\ie it arises from averaging over random
phases), or whether we expect it to hold in a given realization of
disorder, we compute $\ol{|G_{ij}(t,x,x')|^2}$, which is clearly
insensitive to phase fluctuations.  From Eq.~(\ref{eq:GtoGfree}) 
we see that
\beq
\label{eq:|G|^2}
\ol{|G_{ij}(t,x,x')|^2}={1\ov |\XP {\rm n}|^2}
\ol{U_{ij}(x,x')U^{\dag}_{ji}(x,x')}.
\eeq
By the unitarity of $U(x,x')$ we have
\beq
\label{eq:unitarity}
\ol{U_{11}U^{\dag}_{11}}+\ol{U_{12}U^{\dag}_{21}}=1,
\eeq
where we have suppressed the spatial arguments.  Using the explicit
form of $U$ that follows from Eqns.~(\ref{eq:U}), (\ref{eq:Psi,B}),
and $\Im m[\lambda(x)]=0$, one may show $\sigma^y
U^T\sigma^y=U^{\dag}$, which implies
\beq
\label{eq:Utrans}
\ol{U_{11}U^{\dag}_{11}}=\ol{U_{22}U^{\dag}_{22}},\qquad 
\ol{U_{12}U^{\dag}_{21}}=\ol{U_{21}U^{\dag}_{12}}.
\eeq
Finally, from Eq.~(\ref{eq:DisAvgSzSz}) we have 
\beq
\label{eq:fourth}
\ol{U_{11}U^{\dag}_{11}}-\ol{U_{21}U^{\dag}_{12}}
-\ol{U_{12}U^{\dag}_{21}}+\ol{U_{22}U^{\dag}_{22}}=
2\left[{e^{-2|x-x'|W}}\right]^{zz}
\eeq
Eqns.~(\ref{eq:unitarity})-(\ref{eq:fourth}) are four equations in
four unknown quantities which can be solved to yield
\beq
\ol{U_{11}U^{\dag}_{11}}=\ol{U_{22}U^{\dag}_{22}}=
{1\ov2}\left({1+\left[{e^{-2|x-x'|W}}\right]^{zz}}\right),
\quad
\ol{U_{12}U^{\dag}_{21}}=\ol{U_{21}U^{\dag}_{12}}=
{1\ov 2}\left({1-\left[{e^{-2|x-x'|W}}\right]^{zz}}\right).
\label{eq:DisAvgUU}
\eeq
This result, combined with Eq.~(\ref{eq:|G|^2}) gives the disorder-averaged
absolute-magnitudes of the elements of the single-particle Green's
function
\beqarr
\ol{|G_{ij}(t,x,x')|^2}&=&
{1\ov 2|\XP {\rm n} |^2}\left\{{1+(-1)^{i+j}
\exp\left({-{2|x-x'|\ov v_{\rm n}^2}
(\Delta_{\lambda}+\Delta_{\xi}/4)}\right)}\right.\nonumber\\
&&{}\times\left.{
\left[{\cos\left({2|x-x'|\tilde{\lambda}\ov v_{\rm n}}\right)+
{\Delta_{\xi}\ov4v_{\rm n}\tilde{\lambda}}
\sin\left({2|x-x'|\tilde{\lambda}\ov v_{\rm n}}\right)}\right]
}\right\}.
\label{eq:DisAvg|G|^2,1}
\eeqarr

The structure of this result is interesting.  Each $\ol{|G_{ij}|^2}$
is the sum of two terms, one of which is identical to the square of
the free (\ie $\lambda(x)=\xi_{\rm n}(x)=0$) Green's function, and the
other of which has spatial oscillations at the shifted frequency and
an exponential decay in space from the disorder.  The fact that
$\ol{|G_{ij}|^2}$ has a long-ranged part (\ie a term that decays
algebraically rather than exponentially) indicates that in a given
sample $G_{ij}$ is not exponentially damped, and thus from
Eq.~(\ref{eq:tr_from_G}) we can conclude that the neutral mode portion
of the density-density correlation function is also long-ranged for a
given realization of disorder.  We therefore expect that in a given
sample the density two-point function has the structure of the
disorder-averaged quantity (\ref{eq:DisAvgD}), without the exponential
decay in space of the neutral mode piece.  As a function of $\Delta
x\equiv x-x'$ we expect two peaks at the points $\Delta x=v_{\rm
c,n}t$, with the second peak modulated in space at a frequency that
varies with the local scalar potential [see
Eq.~(\ref{eq:shiftlambda})].

\subsubsection{331 Sequence}

In the analysis at the beginning of this section we determined that
for the 331 sequence ($\bh^2=4$), only disorder in the scalar
potential terms was a non-irrelevant perturbation.  We therefore
consider the Hamiltonian
\beqarr
{\cal H}_D&=&\ix \left[{{1\ov 4\pi}v_{\rm c}\,:\!(\delx \phi_{\rm c})^2\!:\,
+{1\ov 4\pi}v_{\rm n}\,:\!(\delx \phi)^2\!:\, + {\lambda\ov 2\pi a}
\left({e^{i2\phi(x)}+e^{-i2\phi(x)}}\right)}\right.\nonumber\\
&&\left.{{}+\xi_{\rm c}(x){1\ov 2\pi}\delx\phi_{\rm c}(x)+
\xi_{\rm n}(x){1 \ov 2\pi}\delx\phi(x)}\right]\label{eq:H331_w/dirt},
\eeqarr
which, with the help of the fermionization (\ref{eq:fermi_331}) for the
neutral boson and the transformation (\ref{eq:eta_c}) for the charged
boson, can be written
\beq
\label{eq:H331_w/dirt2}
{\cal H}_D=\ix :\!\left[{{1\ov 4\pi}v_{\rm c}(\delx\eta)^2 -{i\ov
2}v_1\chi_1\delx\chi_1-{i\ov 2}v_2\chi_2\delx\chi_2
+i\xi_{\rm n}(x)\chi_1\chi_2}\right]\!:, 
\eeq 
\narrowtext
where we have dropped the constant $\xi_{\rm c}^2$ term.  The charged mode
portion of this Hamiltonian is identical to that of
(\ref{eq:H110_quad}) for the 110 sequence, and hence all results
pertaining to the charged mode can be imported from the previous
discussion.  The Hamiltonian is now quadratic, however the lack
of translation invariance prevents us from employing the method used
in Section~\ref{ss:parallelfield}, where we essentially solved the
special case in which $\xi_{\rm c}(x)$ is independent of $x$.  In
addition we cannot absorb the disorder into the definition of the
field operators via a gauge transformation as we did for the 110
sequence, because the Majorana fields are real and therefore neutral.
However, we can still separate the disorder and quantum expectation
values by explicitly constructing solutions to the Heisenberg
equations of motion.  This procedure involves some technical
subtleties not present in the 110 solution, and is discussed in detail
in Appendix~\ref{sec:331-via-spin}.  In this section we use a
different approach; we find the disorder-averaged correlation
functions of the above Hamiltonian by an {\em exact} summation of the
disorder-averaged perturbation theory.

The chirality of the fermions in the neutral mode part of the
Hamiltonian allows a great simplification in the structure of the
diagrammatic perturbation theory in powers of the disorder potential
$\xi_{\rm n}(x)$.  This was first noted by Chalker and Sondhi in the 
context of a single-particle description of the edge
\cite{chalker-sondhi}.   Consider the matrix Green's function of 
the Majorana fields for the free case, \ie $\xi_{\rm n}(x)=0$:
\widetext
\beq
\label{eq:free_MajGF}
g^{(0)}_{ij}(t,x,x')=-i\la
T\chi_i(t,x)\chi_j(0,x')\ra=\delta_{ij}{1\ov \XP j}.
\eeq
Fourier transforming with respect to time gives
\beqarr
g^{(0)}_{ij}(\omega,x,x')&=&\delta_{ij}\int dt\,e^{i\omega t}
{1\ov \XP j}\nonumber\\
&=&\delta_{ij}{i\ov v_j}e^{i\omega (x-x')/v_j}
\left[{\theta(-\omega)\theta(x'-x)-\theta(\omega)\theta(x-x')
}\right].\label{eq:free_MajGF2}
\eeqarr
To obtain this result note that the integrand has a pole in the
complex $t$ plane at $t=(x-x')/v_j+{\cal O}(\epsilon)$.  Therefore for
$x-x'>0$, $\Re e \,t>0$ at the pole and hence the pole lies in the
upper half plane while for $x-x'<0$, $\Re e\,t<0$ at the pole and it
is therefore in the lower half plane.  We find that for positive
frequencies ($\omega>0$) the function vanishes for $x-x'<0$, while for
negative frequencies ($\omega<0$) it vanishes for $x-x'>0$.

Next consider the single-particle Majorana Green's function with the
disorder potential present in Eq.~(\ref{eq:H331_w/dirt2}).  Working
perturbatively in powers of $\xi_{\rm n}(x)$ we have
\beqarr
g_{11}(\omega,x,x')&=&\sum_{n=0}^{\8}\!\int dy_1\ldots dy_{2n}
g^{(0)}_{11}(\omega,x,y_1)\xi_{\rm n}(y_1)g^{(0)}_{22}(\omega,y_1,y_2)
\xi_{\rm n}(y_2)\!\ldots g^{(0)}_{11}(\omega,y_{2n},x'),
\nonumber\\
g_{12}(\omega,x,x')&=&i\sum_{n=0}^{\8}\int\!dy_1\ldots dy_{2n+1}
g^{(0)}_{11}(\omega,x,y_1)\xi_{\rm n}(y_1)
\ldots g^{(0)}_{22}(\omega,y_{2n+1},x'),
\label{eq:MajGF_pert}
\eeqarr
with similar expressions for the remaining components.  Each time the
particle scatters off the impurity potential its velocity changes from
$v_1$ to $v_2$ or vice versa.  When we disorder average the above
equations, we must tie together insertions of $\xi_{\rm n}(x)$ in all
possible ways for each term in the sum.  We first observe that because
the scalar potential is time-independent, $\omega$ is conserved and
hence it has the same sign in every propagator.  This, along with the
chirality of $g^{(0)}(\omega,x,x')$ evident in (\ref{eq:free_MajGF2})
implies that any disorder-averaged diagram in which impurity lines
cross vanishes identically.  Therefore, for each term in the sum in
(\ref{eq:MajGF_pert}) there is a single nonzero disorder-averaged
diagram, \ie the one in which successive insertions of $\xi_{\rm n}$
are pairwise contracted.  The resulting series can be summed to give
\beq
\label{eq:DisAvgMajGF}
\ol{g}_{ij}(\omega,x,x')=\delta_{ij}{i\ov v_j}e^{i\omega
(x-x')/v_j}e^{-\Delta_{\xi}|x-x'|/2v_1v_2}
\left[{\theta(-\omega)\theta(x'-x)-\theta(\omega)\theta(x-x')
}\right].
\eeq
\narrowtext\noindent
We see that the disorder-averaged Green's function retains the chiral
structure of the free Green's function.  While the exponential decay
of the function involves the geometric mean of the two Majorana
velocities, $\sqrt{v_1v_2}$, the frequency dependence only contains
$v_{1(2)}$ for $\ol{g}_{11(22)}$.  This is a direct consequence of the
fact that all terms with crossed impurity lines vanish when we
disorder-average the chiral Green's function.  Therefore, for the
averaged single-particle Green's function, if the particle begins
propagating with velocity $v_1$ it never propagates with velocity
$v_2$, the other velocity enters only through the density of states
when scattering off the potential.  Transforming back to the time
domain we find
\beq
\label{eq:DisAvgMajGF_t}
\ol{g}_{ij}(t,x,x')=\delta_{ij}{e^{-\Delta_{\xi}|x-x'|/2v_1v_2}\ov \XP j}.
\eeq

{}From the relation between the Dirac and Majorana fields,
$\psi=(\chi_1+i\chi_2)/\sqrt{2}$, we find that the Green's function of
the neutral mode fermion, ${\cal G}_{\psi}$ (\ref{eq:prop_331_1}), can
be obtained from $g$ via a unitary transformation
\beq
\label{eq:gtoGpsi}
{\cal G}_{\psi}(t,x,x')=Og(t,x,x')O^{\dag},\;{\rm where}\;\,
O\equiv{1\ov\sqrt{2}}\pmatrix{1 & i \cr 1 & -i \cr}.
\eeq
Thus from Eqns.~(\ref{eq:prop_331_0}), (\ref{eq:DisAvgMajGF}), and
(\ref{eq:gtoGpsi}) we find, after Fourier transforming, that the
single-electron Green's function for the 331 sequence in the presence
of disorder is:
\widetext
\beq
\label{eq:331Gwdirt}
\ol{{\cal G}}(t,x,0)\equiv
{1\over \left[{\X {\rm c}}\right]^{m-1}}{1\ov 2}\left[{
{\one+\sigma^x\ov \X 1}+{\one-\sigma^x\ov \X 2}}\right]
e^{-\Delta_{\xi}|x|/2v_1v_2}.
\eeq
We see that for the single-electron Green's function the velocity
split of the neutral mode remains in the presence of disorder, but the
function acquires an exponential decay with distance.

We next consider the calculation of the density-density correlation
function.  As a first step toward understanding the behavior of this
correlation function in a given sample, we will calculate its
disorder-average.  We can use the transformation (\ref{eq:eta_c}) and the
expression for the Fermi density operator in terms of the Majorana
fields, $:\!\psi^{\dag}\psi\!:= :\!i\chi_1\chi_2\!:$, to write the
density operators (\ref{eq:rho331}) as
\beq
\label{eq:rho331w/dis}
\rho_{1,2}={1\ov 4\pi\sqrt{m-1}}\left({\delx\eta-{1\ov v_{\rm
c}}\xi_{\rm c}}\right)\mp{i\ov 2}\,:\!\chi_1\chi_2\!:,
\eeq
{}From which we find
\beq
\label{eq:D_331}
-i{\cal D}(t,x,x')={1\ov 4(m-1)}{(\one+\sigma^x)\ov [\XP {\rm c}]^2}
-i{(\one-\sigma^x)\ov 4}D(t,x,x'),
\eeq
where we have denoted the neutral mode contribution by
\beq
\label{eq:Dn_331}
-iD(t,x,x')\equiv
\la T\,:\!\chi_1(t,x)\chi_2(t,x)\!:\,:\!\chi_1(0,x')\chi_2(0,x')\!:\,\ra.
\eeq

To evaluate $D$, we first Fourier transform with respect to time in
order to exploit the chirality in the mixed frequency-space domain.
Computing the expectation value in Eq.~(\ref{eq:Dn_331}) using Wick's
theorem and taking the disorder-average then gives:
\beq
\label{eq:DfromF}
-i\ol{D}(\omega,x,x')=-i\int dt\,e^{i\omega t}\ol{D}(t,x,x')\equiv
\int {d\omega'\ov 4\pi}
F\left({{\omega'+\omega\ov 2},{\omega'-\omega\ov 2},x,x'}\right),
\eeq
where we have defined
\beq
\label{eq:Fdef}
F(\omega_1,\omega_2,x,x')=\ol{g_{11}(\omega_1,x,x')g_{22}(-\omega_2,x,x')} 
-\ol{g_{12}(\omega_1,x,x')g_{21}(-\omega_2,x,x')}.
\eeq
If one uses Eq.~(\ref{eq:MajGF_pert}) to write the single-particle
Green's functions in the above expression in terms of the free Green's
function, $g^{(0)}_{ij}$, and the disorder potential, $\xi_{\rm n}$,
one finds upon disorder averaging that the chirality of
Eq.~(\ref{eq:free_MajGF2}) implies that all non-vanishing diagrams are
of the form of ladder diagrams with the legs of the ladder constructed
out of the disorder-averaged propagators $\ol{g}_{ij}$.  

\begin{figure}[htbp]
\begin{center}
\leavevmode
\epsfxsize=\columnwidth
\epsfbox{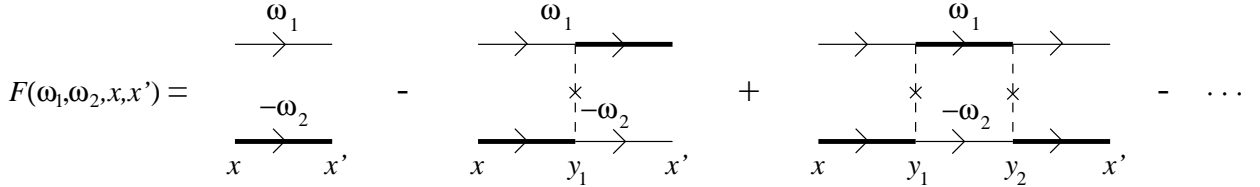}\vskip1mm
\caption{Ladder sum for the neutral mode density-density correlation
function.  The thin solid lines represent $\ol{g}_{11}$, the thick
solid lines represent $\ol{g}_{22}$ and the dashed lines represent the
disorder potential and carry a factor of $\Delta_{\xi}$.}
\label{fig:ladder}
\end{center}
\end{figure}

The legs of the ladder are given by 
\beqarr
\lefteqn{h(\omega_1,\omega_2,k)\equiv\int dx\,e^{-ikx}       
\ol{g}_{11}(\omega_1,x,0)\ol{g}_{22}(-\omega_2,x,0)}&&\nonumber\\
&&={i\ov v_1v_2}\!\left[{{\theta(\omega_1)\theta(-\omega_2)\ov
k-\omega_1/v_1+\omega_2/v_2-i\Delta_{\xi}/v_1v_2}-
{\theta(-\omega_1)\theta(\omega_2)\ov
k-\omega_1/v_1+\omega_2/v_2+i\Delta_{\xi}/v_1v_2}
}\right]\label{eq:h},
\eeqarr
which was evaluated with the help of equation (\ref{eq:DisAvgMajGF})
for $\ol{g}_{ij}$.  The segments of the ladder shown in Fig.~\ref{fig:ladder}
alternate between $h(\omega_1,\omega_2,k)$ and
$h(-\omega_2,-\omega_1,k)$.  Performing the ladder sum gives
\beq
\label{eq:Ffromh}
F(\omega_1,\omega_2,x,x')=\int {dk\ov 2\pi}\,e^{ik(x-x')}\left[{
h(\omega_1,\omega_2,k)-\Delta_{\xi}
h(\omega_1,\omega_2,k)h(-\omega_2,-\omega_1,k)\ov
1-\Delta_{\xi}^2h(\omega_1,\omega_2,k)h(-\omega_2,-\omega_1,k)}\right].
\eeq

Using the result for $h(\omega_1,\omega_2,k)$ in
this expression for $F(\omega_1,\omega_2,x,x')$ gives, after some
algebra,
\beqarr
F(\omega_1,\omega_2,x,x')&=&{i\ov v_1v_2}\int {dk\ov 2\pi}\,e^{ik(x-x')}
\left[{\theta(\omega_1)\theta(-\omega_2)
{k-w(\Delta_{\xi})\ov[k-z_+(\Delta_{\xi})][k-z_-(\Delta_{\xi})]}}\right.
\nonumber\\
&&{}\left.{-\theta(-\omega_1)\theta(\omega_2)
{k-w(-\Delta_{\xi})\ov[k-z_+(-\Delta_{\xi})][k-z_-(-\Delta_{\xi})]}
}\right],\label{eq:Ffromz}
\eeqarr
where we have defined the parameters 
\beqarr
w(\Delta_{\xi})&\equiv&\omega_1/v_2-\omega_2/v_1+2i\Delta_{\xi}/v_1v_2,
\label{eq:w}\\
z_{\pm}(\Delta_{\xi})&\equiv&{1\ov v_1v_2}\left[{
{v_1+v_2\ov 2}(\omega_1-\omega_2)+i\Delta_{\xi}
\pm i\sqrt{\Delta_{\xi}^2-{(v_1-v_2)^2\ov 4}(\omega_1+\omega_2)^2
}}\right].
\label{eq:z}
\eeqarr 
{}From the expression for $z_{\pm}(\Delta_{\xi})$ we see that in the first
term in Eq.~(\ref{eq:Ffromz}) both poles are in the upper half
plane while in the second term both poles are in the lower half plane.
Performing the $k$-integration by the residue theorem and using the
resulting expression for $F$ in Eq.~(\ref{eq:DfromF}) gives:
\beqarr
-i\ol{D}(\omega,x,x')&=&{1\ov v_1v_2}[\theta(-\omega)\theta(x'-x)-
\theta(\omega)\theta(x-x')]
e^{iv_{\rm n}\omega X-\Delta_{\xi}|X|}\nonumber\\
&&\int_{-\omega}^{\omega}{d\omega'\ov 4\pi}
\left[{\cosh(\Lambda(\omega')X)+{\Delta_{\xi}\ov \Lambda(\omega')}
\sinh(\Lambda(\omega')|X|)}\right],\label{eq:DisAvgDomega}
\eeqarr
where we have used the rescaled coordinate $X=(x-x')/v_1v_2$,
and defined
\beq
\label{eq:Lambda}
\Lambda(\omega)\equiv\sqrt{\Delta_{\xi}^2-{1\ov 4}(v_1-v_2)^2\omega^2}
=\sqrt{\Delta_{\xi}^2-{\lambda^2\ov \pi^2}\omega^2}.
\eeq

The experimentally measurable quantity is the retarded
density-density correlation function.  Since
Eq.~(\ref{eq:DisAvgDomega}) is not in the time domain,
we cannot use Eqns.~(\ref{eq:CTO}) and (\ref{eq:CR}) to obtain
$\ol{D^R}(t,x,x')$ directly.  However, in $\omega$-$k$ space we have the
relations
\beq
\label{eq:omega-k}
\Re e\,\ol{D^R}(\omega,k)=\Re e\,\ol{D}(\omega,k),\qquad
\Im m\,\ol{D^R}(\omega,k)=\sgn(\omega)\,\Im m\,\ol{D}(\omega,k),
\eeq
which in turn yield
\beq
\label{eq:omega-x}
\ol{D^R}(\omega,x)=\theta(\omega)\ol{D}(\omega,x)
+\theta(-\omega)[\ol{D}(\omega,-x)]^*,
\eeq
where the asterisk denotes complex conjugation.  Using
Eqns.~(\ref{eq:DisAvgDomega}) and (\ref{eq:omega-x}) and Fourier
transforming back to the time domain we find:
\beqarr
\ol{D^R}(t,x,x')&=&{\theta(x-x')\ov 2\pi}e^{-\Delta_{\xi}X}
{\cal P}\left({1\ov v_{\rm n}X-t}\right)\nonumber\\
&&{}\times \int_{-\8}^{\8}{d\omega\ov 2\pi}
e^{i\omega(v_{\rm n}X-t)} 
\left[{\cosh(\Lambda(\omega)X)+{\Delta_{\xi}\ov \Lambda(\omega)}
\sinh(\Lambda(\omega)X)}\right],
\label{eq:DisAvgDR331,int}
\eeqarr
where ${\cal P}$ denotes the principal value.  The remaining integral
is computed in Appendix~\ref{sec:integrals}, and from
Eqns.~(\ref{eq:sh/L}) and (\ref{eq:ch}) we find
\beqarr
\ol{D^R}(t,x,0)&=&{\theta(t)\ov 2\pi}e^{-\Delta_{\xi}x/v_1v_2}
\left[{{1\ov(v_1-v_2)x}\left\{{v_1\delta(x-v_1t)-v_2\delta(x-v_2t)}\right\}
+{\Delta_{\xi}\ov 2v_1v_2}\theta(z)}\right.\nonumber\\
&&\left.{{\cal P}\left({1\ov v_{\rm n}x/v_1v_2-t}\right)
\left({{x/v_1v_2\ov\sqrt{z}}
I_1\left[{{\Delta_{\xi}\ov\lambda/\pi}\sqrt{z}}\right]
+{\pi\ov \lambda}
I_0\left[{{\Delta_{\xi}\ov\lambda/\pi}\sqrt{z}}\right] 
}\right)}\right],\label{eq:DisAvgDR331}
\eeqarr
\narrowtext\noindent
where $z\equiv (t-x/v_1)(x/v_2-t)$, and $I_n$ are Bessel functions of
imaginary argument.  Note that as expected this function is real.
This result can also be found using the formalism in
Appendix~\ref{sec:331-via-spin}.

We have computed $\ol{D^R}$, but what we are really interested in is
the behavior of $D^R$ in a given realization of disorder.  A
surprising feature of $\ol{D^R}$ is that it has a (principal value)
singularity at the mean arrival time $t_0=v_{\rm n}x/v_1v_2$.  An
immediate question is whether this singularity is present in each
sample, and if not, how does it arise in the average.

To investigate the behavior of $D^R$ in a given realization of
disorder we have adopted several approaches.  First, as in our
analysis of the 110 sequence, we use the fact that $D^R$ can be
expressed in terms of single-particle Green's functions, whose second
moments we evaluate.  Second, we show that in a given sample
$D^R(t,x)$ exhibits an exact symmetry about the point $t=t_0$.
Finally we shall consider the behavior of the correlation functions in
some simple model potentials.

The relation between the time-ordered correlation functions $D$ and
$g_{ij}$ follows from the definition (\ref{eq:Dn_331}):
\beq
\label{eq:Dfromg}
D(t,x,x')=i\det\,g(t,x,x').
\eeq 
Using Eq.~(\ref{eq:CTO}), we then conclude from the above equation
that $D^>=i\det\,g^>$ and $D^<=i\det\,g^<$.  Thus from
Eq.~(\ref{eq:CR}) we have
\beq
\label{eq:DRfromg1}
D^R(t,x,x')=i\theta(t)[\det\, g^>(t,x,x')-\det\, g^<(t,x,x')].
\eeq
If we define the advanced correlation functions $g^A=\theta(-t)(g^<-g^>)$,
then one can show $g^<=g-g^R$ and $g^>=g-g^A$, which when substituted
into Eq.~(\ref{eq:DRfromg1}) give
\beqarr
\label{eq:DRfromg}
D^R&=&i\theta(t)\bigl[(g_{11}g^R_{22}-g_{12}g^R_{21})\nonumber\\
&&+(g^R_{11}g_{22}-g^R_{12}g_{21})
-(g^R_{11}g^R_{22}-g^R_{12}g^R_{21})\bigr].
\eeqarr
We see that the retarded density-density correlation function in a
given sample can be expressed in terms of the time-ordered and
retarded single-particle Green's functions.

We have already evaluated the average time-ordered single-particle
Green's function, see Eq.~(\ref{eq:DisAvgMajGF_t}).  Using
Eqns.~(\ref{eq:CTO}) and (\ref{eq:CR}) we find for the corresponding
retarded function:
\beq
\label{eq:DisAvg_gR}
\ol{g^R_{ij}}(t,x,x')=-i\delta_{ij}\theta(t)\delta(x-x'-v_it)
e^{-\Delta_{\xi}(x-x')/2v_1v_2}.  
\eeq
Next consider the absolute squares of the single-particle Green's
functions.  The disorder average of
$g_{ij}(t,x,x')g^{*}_{ij}(t',x,x')$ can be evaluated by the same
diagrammatic procedure used to obtain the average density-density
correlation function.  The disorder average of
$g^R_{ij}(t,x,x')g^{R*}_{ij}(t',x,x')$ can be readily computed using
the formalism presented in Appendix~\ref{sec:331-via-spin}.

\widetext
Omitting the details of these calculations we find:
\beqarr
\ol{g_{ij}(t,x,0)g^{*}_{ij}(t',x,0)}={1\ov 2\pi}
{e^{-\Delta_{\xi}|X|}\ov \epsilon +i\Delta t}\int_{-\8}^{\8}
{d\omega\ov 2\pi} 
e^{i\omega(v_{\rm n}X-\bar{t})-(\epsilon+i\Delta
t)|\omega|/2}\nonumber\\
\left[{{\delta_{ij}\ov v_j^2} 
\cosh(\Lambda(\omega)X)+\left({{\sigma^z_{ij}(-i\omega\lambda/\pi)
+\sigma^x_{ij}\Delta_{\xi}\ov v_iv_j\Lambda(\omega)}}\right)
\sinh(\Lambda(\omega)X)}\right],\label{eq:gijgij}
\eeqarr
where $\bar{t}\equiv(t+t')/2$ is the average time, $\Delta t\equiv
t-t'$ is the time difference, and $\epsilon >0$ is an infinitesimal
regulator.  For the retarded functions we find
\beqarr
\ol{g^R_{11}(t,x,0)g^{R*}_{11}(t',x,0)}&&\nonumber\\
&&\hskip-1.0in={\theta(t)\ov v_1^2}\delta(\Delta t)e^{-\Delta_{\xi}|X|}
\Biggl[\delta(x/v_1-t)\left.{{}+{\pi\Delta_{\xi}\ov 2\lambda}
\theta(z)I_1\left({{\Delta_{\xi}\ov\lambda/\pi}\sqrt{z}}\right)
\sqrt{x/v_2-t\ov t-x/v_1}}\right],\nonumber\\
\ol{g^R_{12}(t,x,0)g^{R*}_{12}(t',x,0)}&=&
{\theta(t)\ov v_1v_2}\delta(\Delta t)e^{-\Delta_{\xi}|X|}
{\pi\Delta_{\xi}\ov 2\lambda}
\theta(z)I_0\left[{{\Delta_{\xi}\ov\lambda/\pi}\sqrt{z}}\right],
\label{eq:gR2}
\eeqarr
\narrowtext\noindent
where we have again used $z=(t-x/v_1)(x/v_2-t)$.  Note
$\ol{g^R_{21}g^{R*}_{21}}=\ol{g^R_{12}g^{R*}_{12}}$, and
$\ol{g^R_{22}g^{R*}_{22}}$ can be obtained from
$\ol{g^R_{11}g^{R*}_{11}}$ by interchanging $v_1$ and $v_2$.

The first thing to observe is the dependence of these quantities on
the time difference $\Delta t=t-t'$.  In the time-ordered case
(\ref{eq:gijgij}) the dependence is approximately $1/\Delta t$, while
in the retarded case (\ref{eq:gR2}) it is $\delta(\Delta t)$.  This
suggests that in a given configuration both $g$ and $g^R$ are rapidly
varying functions of time.  The remaining integral in the time-ordered
case (\ref{eq:gijgij}) can be evaluated for the special case of equal
times $\Delta t=0$, see Eqns.~(\ref{eq:sh/L}), (\ref{eq:ch}), and
(\ref{eq:omega_sh/L}).  One finds that $\ol{|g_{ij}(t,x,0)|^2}$ is the
same as $\ol{|g^R_{ij}(t,x,0)|^2}$, provided one makes the replacement
$\theta(t)\delta(0)\mapsto 1/2\pi\epsilon$.  Comparing
Eqns.~(\ref{eq:DisAvgDR331}) and (\ref{eq:gR2}) we see that the
structure of the equal-time expressions, $\ol{|g_{ij}(t,x,0)|^2}$ and
$\ol{|g^R_{ij}(t,x,0)|^2}$, is similar to the result for
$\ol{D^R}(t,x,0)$, up to an infinite prefactor.  In particular, the
diagonal elements of $\ol{|g|^2}$ and $\ol{|g^R|^2}$ have
delta-functions with exponentially decaying amplitudes, and all
elements have a term which decays algebraically at large $x$.

In comparing the expressions for $\ol{D^R}$ (\ref{eq:DisAvgDR331}) and
$\ol{|g^R|^2}$ (\ref{eq:gR2}), one obvious difference is the presence
of the factor ${\cal P}[1/(t_0-t)]$ in the density-density correlation
function.  One consequence of this factor is that it makes
$\ol{D^R(t,x,0)}$ at fixed $x$ an odd function about $t=t_0$.  We
shall now demonstrate that this antisymmetry is present in {\rm each}
realization of disorder, not just in the averaged quantity. 

The derivation of this antisymmetry relies on some of the results
derived in Appendix~\ref{sec:331-via-spin} that allow us to express
the Green's function for the disordered 331 sequence in terms of a
coordinate-ordered exponential.  We begin with an expression for the
time-ordered, single-particle matrix Green's function in the presence
of an arbitrary scalar potential $\xi(x)$, (\ref{eq:GPSIF})
\widetext
\beq
\label{eq:GfromAppA}
{\cal G}_{\psi}(t,x,x')=\int {d\omega\over 2\pi i} e^{-i\omega t}\,
\left[\theta(t)\,n(-\omega)-\theta(-t)\,n(\omega)\right]
Q^{-1/2}\,S(x,x';\,\omega)\,  Q^{-1/2},
\eeq
where $n(\omega)\equiv[\exp(\beta\omega)+1]^{-1}$ is the usual Fermi
distribution function (with $\beta$ the inverse temperature), and from
Eqns.~(\ref{eq:Q,f}) and (\ref{eq:defS}):  $Q=v_{\rm n}\one
+(\lambda/\pi)\sigma^x$ and
\beq
\label{eq:SfromAppA}
S(x,x';\,\omega)=T_y\exp\left({i\int_{x'}^{x}dy\, \left[\omega
Q^{-1}-{\xi(y)\ov\sqrt{v_1v_2}}\sigma^z\right]}\right).
\eeq
Eq.~(\ref{eq:GfromAppA}) is analogous to Eq.~(\ref{eq:GtoGfree}) for the 110
sequence.  Both equations express the Green's function in terms of a
coordinate-ordered exponential.  The expression for the 331 sequence
is more complicated because the term in Eq.~(\ref{eq:SfromAppA}) describing
propagation in the absence of disorder, $\omega Q^{-1}$, 
does not commute with the random field term, $\xi(y)\sigma^z$.

Since $Q^{-1}\!=\!(1/v_1v_2)(v_{\rm n}\one-(\lambda/\pi)\sigma^x)$, we can
factor out an overall phase from $S(x,x';\,\omega)$:
\beq
\label{eq:s}
S(x,x';\,\omega)=e^{i\omega v_{\rm n}(x-x')/v_1v_2}
T_y\exp\left({-i\int_{x'}^{x}dy\, \left[{\omega\lambda\ov \pi v_1v_2}
\sigma^x+{\xi(y)\ov\sqrt{v_1v_2}}\sigma^z\right]}\right)
\equiv e^{i\omega t_0}s(x,x';\,\omega).
 \eeq
Using this result and Eq.~(\ref{eq:gtoGpsi}) to transform $G_{\psi}$ to $g$
we find
\beq
\label{eq:g_from_s}
g(t,x,x')={[Q^{-1/2}O]}^{\dag}\int {d\omega\over2\pi i}e^{i\omega(t_0-t)} 
[\theta(t)n(-\omega)-\theta(-t)n(\omega)]s(x,x';\,\omega)[Q^{-1/2}O].
\eeq
{}From this equation we find
\beq
\label{eq:g+*}
g^*(t_0+t,x,x')\!=
\!-{[Q^{-1/2}O]}^T\!\!\!\int {d\omega\over2\pi i}e^{i\omega t}
[\theta(t_0+t)n(-\omega)\!-\!
\theta(-t_0-t)n(\omega)]s^*(x,x';\omega){[Q^{-1/2}O]}^*.
\eeq
{}From the form of the matrix $s$ (\ref{eq:s}), one can show:
\beq
\label{eq:sym_s}
s^*(x,x';\,\omega)=\sigma^y s(x,x';\,\omega)\sigma^y.
\eeq
Combining this with the fact that for $|t|<|t_0|$: 
\beq
\label{eq:thetas}
\theta(t_0+t)n(-\omega)-\theta(-t_0-t)n(\omega)=
\theta(t_0-t)n(-\omega)-\theta(-t_0+t)n(\omega),
\eeq
implies that Eq.~(\ref{eq:g+*}) can be rewritten
\beqarr
&&\!\!\!\!\!\!g^*(t_0+t,x,x')\nonumber\\
&&=-{[Q^{-1/2}O]}^T\sigma^y\int {d\omega\over2\pi i}e^{i\omega t}
[\theta(t_0-t)n(-\omega)-\theta(-t_0+t)n(\omega)]s(x,x';\,\omega)\sigma^y{[Q^{-1/2}O]}^*
\nonumber\\
&&=-{[Q^{-1/2}O]}^T\sigma^y\left({{[Q^{-1/2}O]}^{\dag}}\right)^{-1}
g(t_0-t,x,x'){[Q^{-1/2}O]}^{-1}\sigma^y{[Q^{-1/2}O]}^{*}.
\label{eq:g+tog-}
\eeqarr
Therefore, if we define the matrix
\beq
\label{eq:C}
C\equiv{[Q^{-1/2}O]}^T\sigma^y\left({{[Q^{-1/2}O]}^{\dag}}\right)^{-1}
=\pmatrix{0 & -\sqrt{v_2/v_1} \cr \sqrt{v_1/v_2} & 0},
\eeq
we have the final result that
\beq
\label{eq:sym_g}
g(t_0+t,x,x')=-Cg^*(t_0-t,x,x')C^T,\quad{\rm for}\quad |t|<|t_0|.
\eeq
One can show by an analogous derivation that the corresponding
retarded function, $g^R(t,x,x')$, obeys exactly the same relation.  

Using the relation between $D$ and $g$ (\ref{eq:Dfromg}) and the fact that
$\det\,C=1$, Eq.~(\ref{eq:sym_g}) implies for $|t|<|t_0|$
\beq 
\label{eq:sym_D}
\Re e\,D(t_0+t,x,x')=-\Re e\,D(t_0-t,x,x'),\quad
\Im m\,D(t_0+t,x,x')=\Im m\,D(t_0-t,x,x').
\eeq
Similarly, using the expression for $D^R$ in terms of $g$ and $g^R$
(\ref{eq:DRfromg}), Eq.~(\ref{eq:sym_g}) and the corresponding relation for
$g^R$ gives
\beq
\label{eq:sym_DR}
D^R(t_0+t,x,x')=-D^R(t_0-t,x,x'),
\eeq
where we can drop the restriction $|t|<t_0$ since for times $t$
outside the interval $[0,2t_0]$, $D^R$ is identically zero from
Eq.~(\ref{eq:DisAvgDR331}).  We have found that the retarded
density-density correlation function in a given sample is
antisymmetric about the mean arrival time, indicating that the point
$t=t_0$ is special, independent of $\xi_{\rm n}(x)$.
This result is true at any temperature.  
If we include disorder in the velocity so that $v_{\rm n}$ is a
function of $x$, the symmetry (\ref{eq:sym_DR}) would be absent and
the principal value singularity in $\ol{D^R}$ would be rounded out,
{\em cf.\/}\ Appendix~\ref{sec:random-tunneling}.

\begin{figure}[htbp]
\begin{center}
\leavevmode
\epsfxsize=0.8\columnwidth
\epsfbox{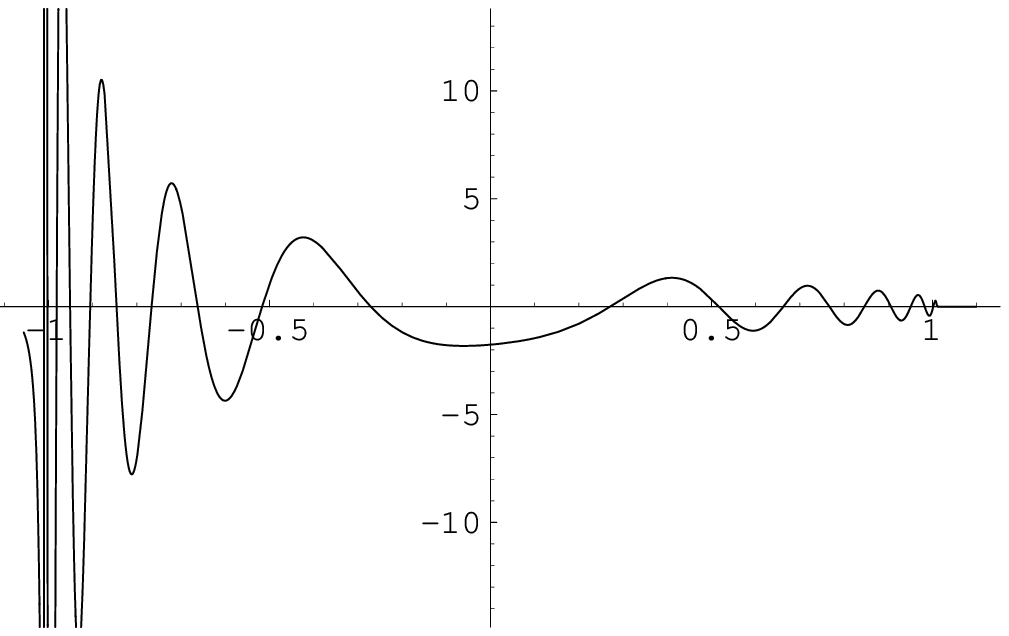}\vskip1mm
\caption{
The imaginary part of $g^R_{11}$ plotted for $v_{\rm n}=1$, $x=10$,
$\lambda/\pi=0.1$, and $\Gamma=3$.  The horizontal axis is time measured
{}from the mean arrival time.
}
\label{fig:g11}
\end{center}
\end{figure}

\begin{figure}[htbp]
\begin{center}
\leavevmode
\epsfxsize=0.8\columnwidth
\epsfbox{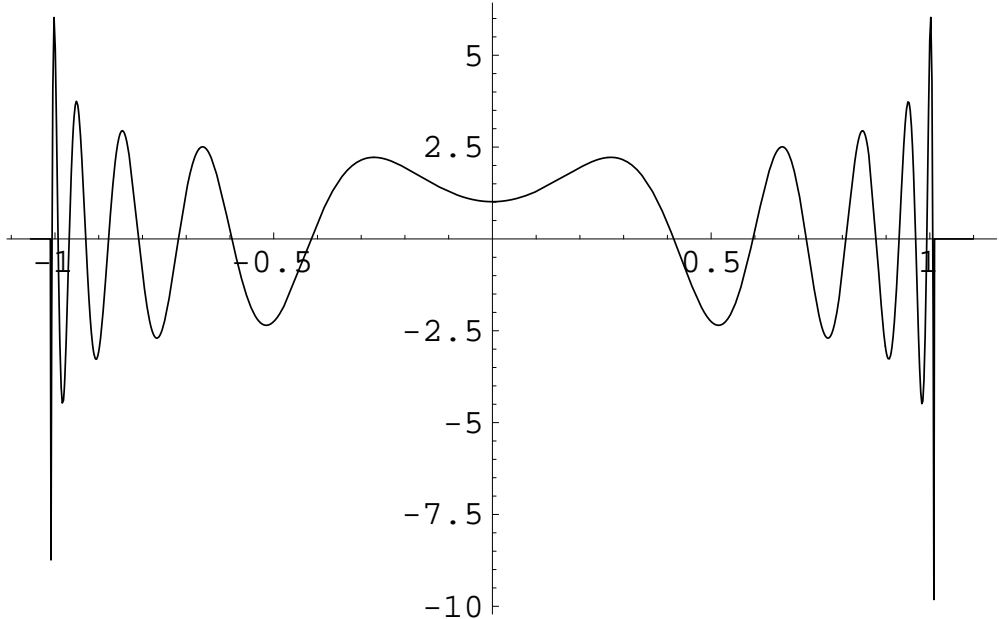}\vskip1mm
\caption{
The imaginary part of $g^R_{12}$ plotted for $v_{\rm n}=1$,$x=10$,
$\lambda/\pi=0.1$, and $\Gamma=3$.  The horizontal axis is time
measured from the mean arrival time.
}
\label{fig:g12}
\end{center}
\end{figure}

Next we consider the behavior of correlation functions for some simple
potentials $\xi_{\rm n}(x)$.  We first consider the case of a uniform
potential.  As we remarked in Section~\ref{ss:parallelfieldS}, the
correlation functions in the presence of a parallel magnetic field can
be reinterpreted as those in the presence of a uniform
scalar potential difference between the layers $\xi_{\rm
n}(x)=\Gamma$.  Using Eqns.~(\ref{eq:CTO}) and (\ref{eq:CR}) to go
{}from the result in Section~\ref{ss:parallelfield} for the
time-ordered single-particle Green's function, ${\cal G}_{\psi}$
(\ref{eq:G_psi_B}), to the corresponding retarded function, and
Eq.~(\ref{eq:gtoGpsi}) to transform this to the Majorana basis, we
find
\beqarr
g^R(t,x)&=&-{2i\theta(t)\ov v_1v_2}\left[{ 
\left({v_{\rm n}\one-{\lambda\ov \pi}\sigma^z}\right) 
\Re e\,P_X(\tau,X)
+\left({{\lambda\ov\pi}\one-v_{\rm n}\sigma^z}\right)
{\lambda\ov \pi}\Re e\,P_{\tau}(\tau,X)}\right.\nonumber\\
&&{}+iv_1v_2\Gamma\sigma^y \Re e\, P(\tau,X)\biggr]\label{eq:gR_wB1}.
\eeqarr
The quantity $\Re e\, P(\tau,X)$ and its derivatives are evaluated in
Appendix~\ref{sec:integrals}, and from Eqns.~(\ref{eq:sin/k}), (\ref{eq:cos}),
and (\ref{eq:omega_sin/k}) we find
\beqarr
g^R_{11}(t,x)&=&
-i\theta(t)\left[{\delta(x-v_1t)-{\pi\Gamma\ov 2\lambda}
\theta(z)J_1\left({{\Gamma\ov\lambda/\pi}\sqrt{v_1v_2z}}\right)
\sqrt{x-v_2t\ov v_1t-x}}\right]\nonumber\\
g^R_{12}(t,x)&=&
-i\theta(t){\pi\Gamma\ov 2\lambda}
\theta(z)J_0\left[{{\Gamma\ov\lambda/\pi}\sqrt{v_1v_2z}}\right],
\label{eq:gR_wB}
\eeqarr
\narrowtext\noindent
where the $J_n$ are standard Bessel functions.  Note $g^R_{21}=-g^R_{12}$ and
$g^R_{22}$ can be obtained from $g^R_{11}$ by interchanging $v_1$ and
$v_2$.  These functions are plotted in Figs.~\ref{fig:g11} and
\ref{fig:g12}.

We see that the delta-functions present in the diagonal elements of
$g^R$ in the case of zero potential (Eq.~(\ref{eq:DisAvg_gR}) evaluated
at $\Delta_{\xi}=0$) remain in the case of a uniform potential.  In
addition to these delta-functions we find oscillatory terms in all elements of
$g^R(t,x)$ considered as functions of $t$ at fixed $x$.
Since the quantity $z=(t-x/v_1)(x/v_2-t)$ is maximal at
$t=t_0$, we see that the frequency of the oscillations in $g^R_{ij}$ is
minimal at the mean arrival time and increases as we approach the
extremal arrival times.  This feature is evident in
Figs.~\ref{fig:g11} and \ref{fig:g12}. 

Finally, we consider the case of a scalar potential made up of isolated
impurities located at the points $\{y_m\}$ with strengths
$\{q_m\}$,\,\ie we take the potential to be
\beq
\label{eq:sumdeltas}
\xi_{\rm n}(x)=\sum_m q_m\delta(x-y_m).
\eeq
Note that the white-noise potential used in the previous calculations
can be approached by the form given in Eq.~(\ref{eq:sumdeltas}) if we
take the number of impurities to infinity and the $q_m$ to be random
variables.  Using Eq.~(\ref{eq:MajGF_pert}), one can compute $g$ for
this potential by an exact summation of the perturbation expansion.
Once again, it is the chirality of $g^{(0)}$ that makes the calculation
tractable.

If there are $N$ impurities between $x$ and $x'$, then the
nonvanishing terms in the perturbation expansion of $g_{ij}$ are in
one-to-one correspondence with the set of all $N$-tuples of
nonnegative integers.  For example, if $x'<y_1<y_2<\ldots<y_N<x$, then
for $t>0$ the nonvanishing terms correspond to propagation from $x'$
to $y_1$ followed by scattering $n_1$ times off impurity $q_1$,
followed by propagation from $y_1$ to $y_2$ and scattering $n_2$ times
off impurity $q_2$, etc., where $n_m$ are nonnegative integers.  If
$n_m$ is even then the velocity of the particle is unchanged by the
scattering (\ie the internal lines on either side of the impurity
are either both $g^{(0)}_{11}$ or both $g^{(0)}_{22}$), while if $n_m$
is odd then the velocity of the particle is changed by the scattering
(\ie the internal lines on either side of the impurity are
$g^{(0)}_{11}$ and $g^{(0)}_{22}$).  Therefore, from the parity
of each element $n_m$ of the $N$-tuple we know which internal
propagators are $g_{11}^{(0)}$ and which are $g_{22}^{(0)}$, and we
can define a corresponding arrival time $T$.  This arrival time is
given by $T=X_1/v_1+X_2/v_2$, where $X_i$ is the total distance the
particle travels with velocity $v_i$.  From these considerations we
can conclude that the general form of the time-ordered and retarded
Green's function for the case of isolated impurities is:
\beqarr
g_{ij}(t,x,x')&=&\sum_k L^{ij}_k{1\ov 2\pi(T_k^{ij}-t+i\epsilon_t)},
\nonumber\\
g^R_{ij}(t,x,x')&=&-i\theta(t)\sum_k L^{ij}_k\delta(T_k^{ij}-t),
\label{eq:g_isoimp}
\eeqarr
where the arrival times $T_k^{ij}$ depend on the end points $x,x'$ and
the location of all impurities $y_m$ located between the end points,
and the coefficients $L^{ij}_k$ depend on the impurity strengths $q_m$.

In calculating $g_{ij}$ with $N$ impurities between $x'$ and $x$,
these impurities divide the interval $[x',x]$ into $N+1$ segments.
Since we are computing $g_{ij}$, the first segment must be associated
with a factor of $g^{(0)}_{ii}$ and the last segment with a factor of
$g^{(0)}_{jj}$.  Since on the $N-1$ remaining segments we can have
either $g^{(0)}_{11}$ or $g^{(0)}_{22}$ depending on the parity of the
$n_m$, there are in general $2^{N-1}$ arrival times $T_k^{ij}$, and
thus the number of terms in the sum in Eq.~(\ref{eq:g_isoimp}) grows
exponentially with the number of intervening impurities.

For a given arrival time $T^{ij}_k$, the parities of the $n_m$'s
are determined and the corresponding coefficient $L^{ij}_k$ in 
Eq.~(\ref{eq:g_isoimp}) is given by
\beq
\label{eq:Lijk}
L^{ij}_k=c_{ij}\prod_{m=1}^N r_{m,p_m},
\eeq
where $c_{ii}\equiv 1/v_i$, $c_{12}\equiv i/\sqrt{v_1v_2}\equiv
-c_{21}$ are overall coefficients, $p_m=n_m \,{\rm mod}\, 2$ is the parity of
$n_m$, and 
\beqarr
r_{m,0}&\equiv&\sum_{n=0}^{\8}\left({iq_m\ov\sqrt{v_1v_2}}\right)^{2n}
={1\ov 1+q_m^2/v_1v_2},
\nonumber\\
r_{m,1}&\equiv&\sum_{n=0}^{\8}\left({iq_m\ov\sqrt{v_1v_2}}\right)^{2n+1}
={iq_m/\sqrt{v_1v_2}\ov 1+q_m^2/v_1v_2},\label{eq:r}
\eeqarr
where we have assumed $|q_m^2/v_1v_2|<1$ for all $m$.  The coefficient
$r_{m,0(1)}$ is the sum of the amplitudes for scattering off the $m$th
impurity an even (odd) number of times.

For a given realization of the potential in Eq.~(\ref{eq:sumdeltas}),
the Green's functions in Eq.~(\ref{eq:g_isoimp}), considered as
functions of time, have an exponentially large number of
singularities.  However, the disorder-averaged quantities, $\ol{g}$
(\ref{eq:DisAvgMajGF_t}) and $\ol{g^R}$ (\ref{eq:DisAvg_gR}), are very
simple functions of time.  This is readily understood if we
approximate the white-noise potential by taking the impurity 
strengths $q_m$ in Eq.~(\ref{eq:sumdeltas}) to be
independent, identically-distributed random variables with zero means.
Then from Eq.~(\ref{eq:r}) we see $\ol{r_{m,1}}=0$ for all $m$.  Thus
the only nonvanishing $L^{ij}_k$ involves an even number of
scatterings off each impurity, and hence there is only {\em one} singularity,
corresponding to the particle never changing its velocity:
\beq
\label{eq:gavg_isoimp}
\ol{g_{ij}}(t,x,0)=\delta_{ij}{1\ov 2\pi(x-x'-v_jt+i\epsilon_t)}
\left({\ol{r_{m,0}}}\right)^N.
\eeq
{}From Eq.~(\ref{eq:r}) we see that $\ol{r_{m,0}}<1$, and since $N$, the
number of impurities between $x'$ and $x$, is proportional to
$|x-x'|$, the last factor in Eq.~(\ref{eq:gavg_isoimp}) reproduces the
exponential decay present in Eq.~(\ref{eq:DisAvgMajGF_t}).

We can understand several things from the general form of the Green's
functions given in Eq.~(\ref{eq:g_isoimp}).  First note that the fact
that $g^R$ for the case of isolated delta-function impurities is a sum
of delta-functions in time (\ref{eq:g_isoimp}) is consistent with the
$\delta(t-t')$ factor present in $\ol{g^R_{ij}(t,x,0)
g^{R*}_{ij}(t',x,0)}$ (\ref{eq:gR2}).  One can similarly show that the
form of $g$ in Eq.~(\ref{eq:g_isoimp}) is consistent with the
$1/(t-t')$ behavior of $\ol{g_{ij}(t,x,0)g_{ij}^*(t',x,0)}$
(\ref{eq:gijgij}).

Next we suppose that the $N$ impurities between $x'$ and $x$ are
evenly spaced.  In the limit $N\rightarrow \8$ it should not make a
difference.  In this case, Eq.~(\ref{eq:g_isoimp}) still holds but the
quantities $L^{ij}_k$ and $T^{ij}_k$ are determined by a different set
of rules, so we shall distinguish them by a tilde.  The number of
arrival times $\Tt^{ij}_k$ is reduced from $2^{N-1}$ down to $N$.
This is because the interval $|x'-x|$ is now divided into $N+1$
segments of {\em equal} length.  The velocity on the first and last
segments is fixed by the indices on $g_{ij}$ and the $N$ arrival times
can then be specified by the number $k$ of the $N-1$ remaining
segments which have velocity $v_2$.  For example, for $g_{11}$ the
arrival times are
\beq
\label{eq:arrival}
\Tt^{11}_k=\left({{k\ov v_2}+{N+1-k\ov v_1}}\right){|x-x'|\ov N+1},
\eeq
where $k=0,1,\ldots (N-1)$.  The amplitude $\Lt^{ij}_k$ is the sum   
of ${N-1 \choose k}$ of the $L^{ij}_k$'s given in Eq.~(\ref{eq:Lijk}).
In the limit of large $N$ the number of terms which contribute to
$\Lt^{ij}_k$ is a Gaussian distribution in $k$ peaked at $k=(N-1)/2$.
The arrival time corresponding to this amplitude is
\beqarr
\label{eq:mean_arrival}
\Tt^{11}_{(N-1)/2}&=&\left({{N-1\ov 2v_2}+ {N+3\ov 2v_1}}\right)
{|x-x'|\ov N+1}\nonumber\\
&\stackrel{N\rightarrow\8}{\longrightarrow}&
{v_{\rm n}|x-x'|\ov v_1v_2}\,=\,t_0.
\eeqarr
Thus we find that the mean arrival time $t_0$ emerges in this model
potential because the number of terms that contribute to the
singularity at $\Tt^{ij}_k$ is maximal for $\Tt^{ij}_k=t_0$.  

All of the considerations in this section lead us to the conclusion
that $D^R(t,x,x')$ in a typical configuration is given by
$\ol{D^R}(t,x,x')$, Eq.~(\ref{eq:DisAvgDR331}), with the factor of
${\cal P}[1/(t_0-t)]$ replaced by a function $f(t,x,x')$, which is a
rapidly fluctuating function of time, antisymmetric about the point
$t=t_0$, and whose amplitude grows as $t$ approaches $t_0$.  The claim
that the general structure of $D^R$ in a typical configuration is
captured by $\ol{D^R}$ is supported by the relation between $D^R$ and
$g,g^R$, Eq.~(\ref{eq:DRfromg}), and the fact that the second moments
$\ol{|g|^2}$ and $\ol{|g^R|^2}$, Eq.~(\ref{eq:gR2}), have the same
structure as $\ol{D^R}$, without the principal value factor.  We
expect that $f(t,x,x')$ is a rapidly fluctuating function of time
based on the dependence of $\ol{g_{ij}(t,x,0)g_{ij}^*(t',x,0)}$ and
$\ol{g^R_{ij}(t,x,0) g^{R*}_{ij}(t',x,0)}$ on $(t-t')$,
Eqns.~(\ref{eq:gijgij}) and (\ref{eq:gR2}), and the behavior of $g$
and $g^R$ for the case of isolated delta-function impurities,
Eq.~(\ref{eq:g_isoimp}).  However complicated $f(t,x,x')$ is, we know
it must be antisymmetric about $t=t_0$ by Eq.~(\ref{eq:sym_DR}).  The
claim that the amplitude of $f(t,x,x')$ approaches a maximum at
$t=t_0$ is supported by several results.  First, recall that $g^R$ in
a constant potential $\Gamma$, Eq.~(\ref{eq:gR_wB}), is oscillatory
with a frequency which is proportional to $\Gamma$ and which is
minimal at the mean arrival time.  This suggests that in a potential
which varies with $x$ there will be less cancelation near the mean
arrival time than near the extremal arrival times.  Second, there is
the observation that in the model of equally-spaced isolated
impurities the number of terms which contribute to each singularity is
maximal for the singularity at the mean arrival time.  Finally, we
have the fact that averaging produces a function singular at $t=t_0$.
These conclusions are supported by our numerical simulations,
see Section~\ref{sec:summary}.

\subsection{Pfaffian Edge}
\label{ss:pfaffian}

In this section we consider the edge theory of the Pfaffian
state, concentrating on the form the retarded density response
function in the presence of a finite temperature and disorder. 

The edge theory of the Pfaffian state contains a $c=1/2$ minimal model
CFT in addition to the usual ($c=1$) chiral boson\cite{imura-ino}.
Recall that the $c=1/2$ minimal model has three primary fields, which
we denote as $I,\chi$, and $\sigma$, and whose scaling dimensions are
0, 1/2, and 1/16, respectively.  The $\chi$ field is identical to a
single chiral Majorana fermion.  The chiral boson in the edge theory,
$\varphi$, can be shown to have a compactification radius of
$R=1/\sqrt{8}$, the same radius as the charged mode of the 331 state,
see {\bf I}.  In addition to the $c=1/2$
primary fields and their descendents, the operator content of the edge
theory includes the primary field $\delx\varphi$ and the vertex
operators $e^{ik\varphi/\sqrt{8}}$, where $k\in\Zint$.  The electron
creation operator is\cite{milo-read}
\beq
\label{eq:PfaffianElec}
:\!\Psi^{\dag}(x)\!:={1\ov 2\pi a}\chi(x)e^{-i\sqrt{2}\varphi(x)},
\eeq
which has a scaling dimension of 3/2.

We now determine the electric charge density operator, $\rho(x)$, for
the Pfaffian edge.  One requirement for this operator is that the
electron operator has a unit charge with respect to it
\beq
\label{eq:PfCR}
[\rho(x),\Psi^{\dag}(x')]=\delta(x-x')\Psi^{\dag}(x').
\eeq
One candidate which satisfies this relation is
\beq
\label{eq:PfaffianRho}
\rho(x)={1\ov 2\pi\sqrt{2}}\delx\varphi(x).
\eeq
The operator on the r.h.s. of this equation has dimension
one.  The only other dimension one operators present in the edge theory
are $e^{\pm i\sqrt{2}\varphi}$.  However, if we were to add to our
definition of $\rho(x)$ some non-zero multiple of the Hermitian combination
$(e^{i\sqrt{2}\varphi}+e^{-i\sqrt{2}\varphi})$, we would find that the
condition (\ref{eq:PfCR}) is violated.  Any other 
higher-dimension operators which could be added to $\rho(x)$ would
necessarily be multiplied by explicit powers of the short distance
cutoff $a$, and will therefore vanish in the continuum limit.
The Pfaffian state has no interlayer dynamics.  Each electron is in a
state symmetric between the layers, under the assumption that the 
symmetric-antisymmetric splitting is large.  Hence, there is an
energy gap for any process which excites the layers independently.

Having determined the charge density operator for the Pfaffian edge
theory, we can immediately write down the retarded density two-point
correlation function at zero temperature for the clean system
\beq
\label{eq:PfDR}
{\cal D}^R_{\rm Pf}(t,x)=-{1\ov 4\pi}\theta(t)\delta'(x-v_{\varphi}t).
\eeq

{}From the results in Section~\ref{ss:disorder} we know that the only
RG non-irrelevant disorder is scalar potential disorder.  This does
not modify ${\cal D}^R_{\rm Pf}$ at all, just as in the case of
the charged mode for both the 110 and 331 sequences.  From the
discussion in Section~\ref{ss:finitetempS} we know that at a finite
temperature this correlation function is also unchanged.  We find that
${\cal D}^R_{\rm Pf}(t,x)$ has a signal at a single velocity even at a
finite temperature and in the presence of a non-zero scalar potential.

\subsection{Numerics}
\label{ss:numerics}

Here we briefly discuss the numerics for the neutral mode contribution
to the retarded density response function of the 331 edge.  The
matrices $S(x,0;\,\omega)$ were computed using a discretized version of
Eq.~(\ref{eq:SfromAppA}).  Using Eq.~(\ref{eq:GfromAppA}) the
single-particle Green's function was then found by a fast Fourier
transform (FFT) algorithm.  Finally, we used the relation between the
single-particle Green's functions and the density response function
(\ref{eq:DRfromg1}) to calculate $D^R$, and integrated over time to
find the neutral mode contribution to the signal in
Eq.~(\ref{eq:lin_resp2}).

\section{Acknowledgments}
\label{sec:ack}
We would like to thank A. H. MacDonald for suggesting we consider
gated samples, and C. de C. Chamon for bringing Ref.~\cite{wen2} to our
attention.  We would like to acknowledge support by an NSF Graduate
Research Fellowship (JDN), DOE Grant DE-FG02-90ER40542 (LPP) as well
as NSF grant No. DMR-99-78074, US-Israel BSF grant No. 9600294, and
fellowships from the A. P. Sloan Foundation and the David and Lucille
Packard Foundation (SLS).  LPP and SLS thank the Aspen Center for
Physics for its hospitality during the completion of part of this
work.

\appendix

\section{Disordered 331 Sequence via the Spin-analogy}
\label{sec:331-via-spin}

In this appendix we present an alternative method for obtaining exact
disorder-averaged correlation functions for the 331 sequence.  It is
similar to the method used in our discussion of the 110 sequence in
Section~\ref{ss:disorder}, with some technical complications.  The
procedure is based on the fact that we can explicitly construct
solutions to the Heisenberg equations of motion for each realization
of disorder.  We shall ignore the charged mode throughout the
discussion.  

\widetext
The 331 Hamiltonian including a disordered scalar potential
(\ref{eq:H331_w/dirt2}) expressed in terms of the Dirac field is
\beq
\label{eq:HD331D}
{\cal H}_D=\ix:\!\left[{-iv_{\rm n}\psi^{\dag}\delx\psi
-i{\lambda\ov 2\pi}(\psi^{\dag}\delx\psi^{\dag}+\psi\delx\psi)
+\xi(x)\psi^{\dag}\psi}\right]\!:,
\eeq
where in this appendix we suppress the subscript on $\xi$ and take
periodic boundary conditions $\psi(x+L)=\psi(x)$.  The Heisenberg
equations of motion for the field operator $\psi(t,x)$ and its
Hermitian conjugate are:
\beqarr
\left[{\del_t+v_{\rm n}\del_x+i\xi(x)}\right] \psi(t,x)+(\lambda/\pi)\del_x
\psi^{\dag}(t,x)&=&0,\nonumber\\
\left[{\del_t+v_{\rm n}\del_x-i\xi(x)}\right]\psi^{\dag}(t,x)
+(\lambda/\pi)\del_x
\psi(t,x)&=&0\label{eq:HEOM}.
\eeqarr
\narrowtext\noindent
The anomalous (\ie fermion-number non-conserving) terms in ${\cal
H}_D$ couple the equations of motion for $\psi$ and $\psi^{\dag}$ and
therefore we must expand the field in terms of both creation and
annihilation operators
\beq
\label{eq:psifroma}
\psi(t,x)=\sum_n \left[{A_n(x)e^{-i\omega_nt}a_n+ 
B_n(x)e^{i\omega_nt}a_n^{\dag}}\right],
\eeq
where by assumption $a^{\dag}_n$ are canonical Fermi 
operators which create exact single-particle eigenstates of ${\cal
H}_D$ with energies $\omega_n$.  Substituting this expansion into the
equations of motion leads to the following matrix differential equation
\beq
\label{eq:Qdiffeq}
-iQ\delx f_n(x)=[\omega_n\one-\xi(x)\sigma^z]f_n(x),
\eeq
where
\beq
\label{eq:Q,f}
Q\equiv\pmatrix{v_{\rm n} & \lambda/\pi \cr \lambda/\pi & v_{\rm n}},\qquad
f_n(x)\equiv \pmatrix{A_n(x) \cr B^{*}_n(x)}.
\eeq

Note that the Hamiltonian (\ref{eq:HD331D}) has a particle-hole
symmetry under which $\psi\leftrightarrow \psi^{\dag}$ and
$\xi\mapsto-\xi$.  In terms of the two-component wavefunction
$f_n(x)$, this implies that if $f_n$ is a solution to
Eq.~(\ref{eq:Qdiffeq}) with energy $\omega_n$ then
$\tilde{f}_n=\sigma^xf^{*}_n$ is a solution with energy $-\omega_n$.
Assuming all $\omega_n\neq0$, we can enumerate the functions $f_n(x)$
in such a way that $\omega_{-n}=-\omega_n$ and $\omega_n>0$ for $n>0$.
This implies $\tilde{f}_{-n}=f_n$, from which we find $A_n=B_{-n}$, an
indication that some double counting may be present.  Indeed, the
particle--hole symmetry only interchanges the two
equations~(\ref{eq:HEOM}); it should not generate new solutions.  This
double counting can be removed if we define $a_n+a^{\dag}_{-n}\mapsto
a_n$, and write, instead of Eq.~(\ref{eq:psifroma}),
\beq
\label{eq:psifromf}
\pmatrix{\psi(t,x) \cr \psi^{\dag}(t,x)} =\sum_nf_n(x)\,e^{-i\omega_n
  t}\,a_n.  
\eeq
To obtain the solutions to Eq.~(\ref{eq:Qdiffeq}) we define the
rescaled wavefunctions 
\beq 
\label{eq:zeta}
\zeta_n=Q^{1/2}\,f_n,
\eeq
in terms of which the differential equation~(\ref{eq:Qdiffeq}) becomes
\beq
\label{eq:zetaDiffEq}
\delx\zeta_n(x)=i\left[{\omega_nQ^{-1}-{\xi(x)\ov\sqrt{v_1 v_2}}
\sigma^z}\right]\,\zeta_n(x).
\eeq
We can write the solutions to this equation as 
\beq
\label{eq:zetafromS} 
\zeta_n(x)\equiv S(x,0;\,\omega_n)\,\zeta_n(0), 
\eeq 
with a coordinate-ordered exponential 
\beq 
\label{eq:defS}
S(x,x';\,\omega)=T_y\exp\left({i\int_{x'}^{x}dy\, \left[\omega
Q^{-1}-{\xi(y)\ov\sqrt{v_1v_2}}\sigma^z\right]}\right),
\eeq 
for $x>x'$, 
and the Hermitian conjugate $S(x,x';\,\omega)=
S^\dagger(x',x;\,\omega)$ for $x<x'$.  The boundary conditions on the
Fermi field imply $\zeta_n(x+L)=\zeta_n(x)$, which in turn means that
the allowed energies $\omega_n$ are determined by finding those
energies for which the matrix $S(L,0;\,\omega)$ has a unit eigenvalue
with the corresponding eigenvector taken to be $\zeta_n(0)$,
\begin{equation}
  \label{eq:eigenproblem-zeta}
  \left[S(L,0;\,\omega)-{\bf 1}\right]\zeta_n(0)=0.
\end{equation}

The orthogonality of solutions $f_n(x)$ for different values of
$\omega_n$ is guaranteed by the fact that the differential
equation~(\ref{eq:Qdiffeq}) is self-conjugate, while their
normalization must be demanded explicitly,
\beq 
\label{eq:normalf}
\ix f^{\dag}_n(x)f_n(x)=1.
\eeq
This can be rewritten with the help of Eqns.~(\ref{eq:zeta}) and
(\ref{eq:zetafromS}) as 
\beqarr
\label{eq:normalzeta}
&\displaystyle
L{v_{\rm n}\ov v_1v_2}\zeta^{\dag}(0){\cal N}(\omega,\xi)\zeta(0)=1,&\\
\nonumber&  \displaystyle
\lefteqn{{\cal N}(\omega,\xi)=\one-{\lambda\ov \pi
v_{\rm n}L}\ix S^{\dag}(x,0;\,\omega)\sigma^xS(x,0;\,\omega).}\hskip3in&
\eeqarr
Usually it is the wavefunction normalization that makes the disorder
calculations so difficult.  Note, however, that the integration in the
second term of the normalization matrix ${\cal N}(\omega,\xi)$ is
extended over the entire length of the sample, which makes it a
self-averaging object.  As in Section~\ref{ss:disorder}, the matrix
$S(x,x';\,\omega)$ (\ref{eq:defS}) can be interpreted as the evolution
operator in a fictitious time $y$, for a spin precessing under the
influence of a constant magnetic field in the $x$-direction (due to
the off-diagonal terms of the matrix $Q^{-1}$), and a random field
$\sim\xi(x)$ along the $z$-direction.  The integral in
Eq.~(\ref{eq:normalzeta}) is the $x$-component of the spin averaged
over a ``time'' $L$.  In the presence of a non-vanishing disorder, the
initial orientation is forgotten after a finite distance, and the
second term in the normalization matrix~(\ref{eq:normalzeta})
disappears in the thermodynamic limit, $L\rightarrow\8$.  This happens
with probability one for any realization of disorder.  Physically,
this simplification is related to the fact that in a chiral system
localization does not happen; each particle explores the entire
circumference of the sample.

Let us now consider the single-particle Green's function of the
neutral mode fermion, ${\cal G}_{\psi}$ [Eq.~(\ref{eq:prop_331_1})].
Using Eq.~(\ref{eq:psifromf}), this can be written 
\widetext
\beq
\label{eq:GPSI1}
{\cal G}_{\psi}(t,x,x')=-i\sum_n f_n(x)f^{\dag}_n(x')e^{-i\omega_nt}
[\theta(t)\,n(-\omega_n)-\theta(-t)\,n(\omega_n)], 
\eeq
where
$n(\omega)\equiv[\exp(\beta\omega)+1]^{-1}$ is the usual Fermi
distribution function.  Using Eqns.~(\ref{eq:zeta}),
(\ref{eq:zetafromS}), and (\ref{eq:normalzeta}), with ${\cal
  N}(\omega,\xi)=\one$, we obtain 
\beqarr
{\cal G}_{\psi}(t,x,x')&=&-i\,\left({v_1v_2\over
    L\,v_{\rm n}}\right)\sum_n e^{-i\omega_nt}\,
\left[\theta(t)\,n(-\omega_n)-\theta(-t)\,n(\omega_n)\right]
\nonumber\\
&\relax & 
\times Q^{-1/2}\,\left[{
      {S(x,0,\omega_n){\zeta_n(0)
        \zeta_n^{\dag}(0)\ov \zeta_n^{\dag}(0)\zeta_n(0)}
S^{\dag}(x',0,\omega_n)
        }}\right]
Q^{-1/2},\label{eq:GPSI2} 
\eeqarr
where $\zeta_n(0)$ obey the eigenvalue
equation~(\ref{eq:eigenproblem-zeta}).   
To take the thermodynamic limit, we need to set the system size to
infinity, keeping other parameters (temperature, disorder, distance
$|x-x'|$, etc.) finite.  Effectively, this implies that we can select
an energy interval $\Delta E$, ``infinitesimal'' on a scale defined by
these finite quantities, and yet containing a macroscopic number of
energy levels, such that the averaging over the states within this
interval gives

$$
\left\langle \zeta_n\,\zeta_n^\dagger\right\rangle_{\omega_n\in \Delta
  E}= {1\over2}\,{\bf 1}.
$$
The value of the average and the existence of such an interval follows
{}from the fact that for any $\omega_1\neq\omega_2$, the spin-rotation
matrices $S(L,0;\,\omega_{1,2})$ become entirely uncorrelated for a
sufficiently large $L$, or, equivalently, the relative rotation matrix
$
S^\dagger(L,0;\,\omega_1)\,S(L,0;\,\omega_2) 
$
entirely forgets the initial direction. 

Performing the averaging over the eigenstates within such an
interval, we obtain for the correlation function~(\ref{eq:GPSI2}),
\begin{equation}
  {\cal G}_{\psi}(t,x,x')=-i\,\int {d\omega\over2\pi} e^{-i\omega t}\,
  \left[\theta(t)\,n(-\omega)-\theta(-t)\,n(\omega)\right]
  Q^{-1/2}\,S(x,x';\,\omega)\,  Q^{-1/2},\label{eq:GPSIF} 
\end{equation}
where the summation was replaced by an integration using the ``clean''
single-particle total density of states
$\bar{\rho}=(v_1^{-1}+v_2^{-1})L=2L v_{\rm n}/v_1v_2$, which cannot be
modified by disorder.  Note that Eq.~(\ref{eq:GPSIF}) does not contain
a disorder average; it is an expression valid for any given
realization of disorder (or even in the limit of no disorder, as long
as this limit is taken {\em after\/} the thermodynamic limit).  For
example, we checked that Eq.~(\ref{eq:GPSIF}) with $\xi(x)={\rm
const.}$ reproduces Eq.~(\ref{eq:G_psi_B}), which was derived by more
conventional methods.

{}From the definition of $S(x,x';\,\omega)$, Eq.~(\ref{eq:defS}), and
the disorder averaging procedure used previously, see
Eq.~(\ref{eq:DisAvgU}), we find 
\beq
\label{eq:DisAvgS}
\ol{S(x,x';\,\omega)}=e^{-\Delta_{\xi}|x-x'|/2v_1v_2}e^{i\omega(x-x')Q^{-1}}. 
\eeq
With the help of Eqns.~(\ref{eq:prop_331_0}) and (\ref{eq:GPSIF}) this
gives, in the zero temperature limit, 
\beq
\ol{{\cal G}(t,x,0)}\equiv
{1\over \left[{\X {\rm c}}\right]^{m-1}}{1\ov 2}\left[{
{\one+\sigma^x\ov \X 1}+{\one-\sigma^x\ov \X 2}}\right]
e^{-\Delta_{\xi}|x|/2v_1v_2}.
\eeq
This is in exact agreement with Eq.~(\ref{eq:331Gwdirt}) of
Section~\ref{ss:disorder}.  We have checked that the other
disorder-averaged correlation functions for the 331 sequence discussed
in Section~\ref{ss:disorder} can also be reproduced using the method
described in this appendix.

\section{Random Tunneling for the 331 Sequence}
\label{sec:random-tunneling}

In this appendix we illustrate the effect of an RG irrelevant
random perturbation by analyzing the neutral mode of the 331 bilayer
in the presence of velocity and tunneling disorder.  Specifically, we
assume that both the neutral mode velocity $v_{\rm n}(x)$ and the
tunneling amplitude $\lambda(x)$ in Eq.~(\ref{eq:HD331D}) are
coordinate-dependent, in such a fashion that the system remains
chiral, $v_{1,2}(x)=v_{\rm n}(x)\pm\lambda(x)/\pi>0$ for all $x$.  The
introduction of such a coordinate dependence requires only a slight
modification of the Hamiltonian~(\ref{eq:HD331D}).  Specifically, the
first term in the Hamiltonian density must be replaced as
$$
-iv_{\rm n}\psi^\dagger\partial_x\psi\to-{i\over2}\left\{
    v_{\rm n}(x)\,\psi^\dagger\partial_x\psi+\psi^\dagger\partial_x\left[
    v_{\rm n}(x)\psi\right] \right\}.
$$
The arguments in Appendix~\ref{sec:331-via-spin} can then be repeated
with little modification and we obtain, in place of
Eqns.~(\ref{eq:GPSI2}), (\ref{eq:defS}),
\begin{eqnarray}
  {\cal G}_{\psi}(t,x,x')&=&\int {d\omega\over2\pi i} e^{-i\omega t}\,
  \left[\theta(t)\,n(-\omega)-\theta(-t)\,n(\omega)\right]
  Q^{-1/2}(x)\,S(x,x';\,\omega)\,  Q^{-1/2}(x'),\label{eq:GPSIF-x}
  \\
  S(x,x';\,\omega)&=&
  T_y\exp\left({i\int_{x'}^{x}dy\, \left[\omega
    Q^{-1}(y)-u(y)\sigma_z\right]}\right), \quad
  u(y)\equiv {\xi(y) \over {v^{1/2}_1(y)\,v^{1/2}_2(y)}}.
  \label{eq:GF-full331}
\end{eqnarray}
Again, this expression is valid for any given configuration of
disordered $v_{\rm n}(x)$, $\lambda(x)$, and $\xi(x)$.  

\begin{figure}[htbp]
  \begin{center}
    \leavevmode
    \epsfxsize=0.7\columnwidth
    \epsfbox{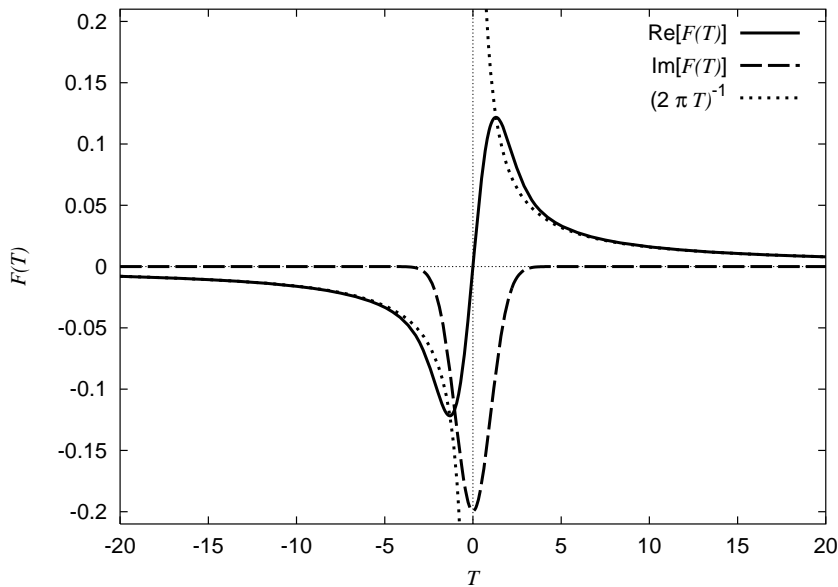}\vskip1mm
    \caption{The real and imaginary parts of the universal function $F(T)$
    [Eq.~(\protect\ref{eq:F-of-T})], which describes the shape of the
    peaks of the averaged Green's function for the 331 double layer.
    Dotted line shows the real part of the Green's function in the absence
    of disorder.}
    \label{fig:error-function}
  \end{center}
\end{figure}

The
requirement $v_{1,2}(x)>0$ is equivalent to essentially non-Gaussian
disorder, and the disorder averaging is generally non-trivial.  This,
however, 
is greatly simplified in the absence of potential disorder,
$\xi(x)=0$.  In this case the remaining matrices $Q^{-1}(y)$ in the
exponential commute with one another for all $y$, the coordinate
ordering ($T_y$) can be omitted, and the disorder averaging can be
performed directly.  The structure of the expression is most evident
after the unitary transformation~(\ref{eq:gtoGpsi}) to the Majorana fermion
representation,
\begin{eqnarray}
  g_{ij}(t,x,x')&=&O^\dagger{\cal G}_\psi(t,x,x')\,O
  \nonumber\\  &=&
  \int {d\omega\over 2\pi i}\,
  e^{-i\omega t}\,
  \left[\theta(t)\,n(-\omega)-\theta(-t)\,n(\omega)\right]\,
  {\delta_{ij}\,\exp[i\omega\tau_i(x,x')]\over
  v_i^{1/2}(x)\,v_i^{1/2}(x')},\label{eq:gf-majorana}
\end{eqnarray}
where $\tau_i(x,x')\equiv \int_{x'}^x dy/v_i(y)$ is the time it takes
for the $i$\,th mode to travel from $x'$ to $x$.  Clearly, the
structure of the correlation function in a given configuration of
disorder does not change; at zero temperature we obtain [{\em cf}.\
Eqns.~(\ref{eq:CSG_G331}), (\ref{eq:331Gwdirt})]
$$
{{\cal G}}_\psi(t,x,x')=
{1\ov 4\pi}\left[{
{\one+\sigma^x\ov v_1^{1/2}(x)\,v_1^{1/2}(x')\,(\tau_1(x,x')-t)}+
{\one-\sigma^x\ov v_2^{1/2}(x)\,v_2^{1/2}(x')\,(\tau_2(x,x')-t)}}\right].
$$
The disorder averaging can be performed for weak disorder if we
notice that the velocity fluctuations along the entire path contribute
to the arrival times $\tau_i(x,x')$; these quantities acquire nearly
Gaussian distributions at sufficiently large distances (compared to the
disorder correlation length, $l_c\ll |x-x'|$).  If we ignore small
multiplicative corrections near the ends of the interval, we then find
$$
\ol{{\cal G}_\psi(t,x,0)}=
{1\ov 4\pi}\left[
  \left(\,\ol{v_1^{-1/2}}\,\right)^2
  {\one+\sigma^x\over D_1^{1/2}}\,F\left(T_1/D_1^{1/2}\right) +
  \left(\,\ol{v_2^{-1/2}}\,\right)^2
  {\one-\sigma^x\over D_2^{1/2}}\,F\left(T_2/D_2^{1/2}\right)
\right],
$$
\narrowtext\noindent
where $T_i=T_i(x)\equiv \bar\tau_i(x)-t$ is the time elapsed from
the arrival of the $i$\,th peak, $D_i=
D_i(x)\equiv\ol{\tau^2(x)}-\ol\tau^2(x)$ is the corresponding
dispersion, and
\begin{equation}
  F(T)\equiv \int_0^\infty {d\omega\over 2\pi i}\,e^{i\omega
    T-\omega^2 /2},
  \label{eq:F-of-T}
\end{equation}
where we have assumed $t>0$.  As illustrated in
Fig.~\ref{fig:error-function}, at large values of the argument, this
function approaches asymptotically the clean single-particle
Green's function, $F(T)=(2\pi T)^{-1}$, $|T|\gg1$.
Although the perturbation is RG irrelevant, the form of the
Green's function is modified in the vicinity of the singularities.

For a weak ($w\ll v_{\rm n}$) Gaussian disorder with a finite correlation
length,
$$
\lambda(x)=\bar\lambda+\delta\lambda(x),\quad
\langle\delta\lambda(x)\,\delta\lambda(y)\rangle=w^2f(x-y)\quad f(0)=1,
$$
assuming $w^2\ll \bar\lambda^2$,
we obtain, to leading order in the weak disorder expansion,
$$
\ol{\tau_i(x)}=x/\bar v\,[1+w^2/\pi^2\bar v_i^2+{\cal O}(w^4/\bar
v_i^4)],
$$
where $ D_i(x)=w^2 x l_c/v_i^4+\ldots$, and the disorder correlation length,
$$
l_c=\int_0^\infty dx \,f(x),
$$
was assumed to be short compared with the overall distance, $ l_c\ll
|x|$.

\section{Evaluation of Integrals}
\label{sec:integrals}

In this appendix we evaluate the integrals needed in the main text.
The basic integral is of the form
\beq
\label{eq:A}
A(\tau,X)=\ii {d\omega\ov 2 \pi}e^{i\omega \tau}
{\sinh(\Lambda(\omega)X)\ov\Lambda(\omega)},
\eeq
where $\Lambda(\omega)=\sqrt{a^2-b^2\omega^2}$, and $\tau, X, a$, and
$b$ are real parameters.  At large frequencies the integrand is of the
form $1/\omega$ times an oscillatory function, and therefore converges
without a regulator.  Also, since the expansion of the integrand in
powers of $\Lambda(\omega)$ contains only even powers, no branch cut
is needed and the integrand is therefore an analytic function for finite
$\omega$.  For large $\omega$ the integrand contains the factors
$e^{i\omega(\tau\pm b X)}$.  Thus for $|bX|<|\tau|$, the integrand is
exponentially small on one side of the real axis and we can therefore
close the contour on that side and find $A=0$, hence
\beq
\label{eq:ThetaFunc}
A(\tau,X)\propto \theta[(b X)^2 - \tau^2].
\eeq

\begin{figure}[htbp]
\begin{center}
\leavevmode
\epsfxsize=0.7\columnwidth
\epsfbox{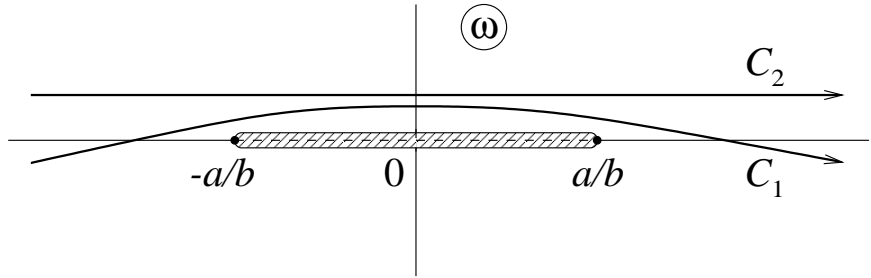}\vskip1mm
\caption{The complex $\omega$ plane showing the branch cut along the real
$\omega$ axis and the contours $C_1$ and $C_2$.}
\label{fig:int}
\end{center}
\end{figure}

To evaluate the integral in the region for which it is nonzero, we
break up the integral into two terms:
\beq
\label{eq:A2}
A(\tau,X)=\int_{C_1}{d\omega\ov 2\pi} 
e^{i\omega\tau}{e^{\Lambda(\omega)X}\ov 2\Lambda(\omega)}-
\int_{C_2}{d\omega\ov 2\pi} e^{i\omega\tau}
{e^{-\Lambda(\omega)X}\ov 2\Lambda(\omega)},
\eeq
where now we must introduce a branch cut, which we take to run along
the real $\omega$ axis from $-a/b$ to $a/b$.  The contours $C_1$ and
$C_2$ are shown in Fig.~\ref{fig:int}.  Assuming $X>0$, the
contours have been chosen so that the integrals in Eq.~(\ref{eq:A2})
are separately convergent.  The integrand in the $C_2$ integral is
exponentially small in the upper half plane and thus can be closed
there to give zero.  The integrand in the $C_1$ integral is
exponentially small in the lower half plane and can be closed there
and contracted to run around the branch cut.  We use the change of
variables $\omega=(a/b)\sin\varphi$ in the integral around the cut to
arrive at
\beq
\label{eq:A3}
A(\tau,X)= \theta[(b X)^2 - \tau^2]{1\ov 2b}\int_0^{2\pi}
{d\varphi\ov 2\pi}e^{a[i(\tau/b)\sin\varphi +X\cos\varphi]}.
\eeq

We next note that we can write
\beq
\label{eq:shift}
i{\tau\ov b}\sin\varphi+X\cos\varphi=\sqrt{X^2-\left({\tau\ov
b}\right)^2}
\cos(\varphi-i\varphi_0),
\eeq
where the real parameter $\varphi_0$ is defined by $\cosh \varphi_0=
X/\sqrt{X^2-(\tau/b)^2}$.  After performing this substitution in
Eq.~(\ref{eq:A3}) and noting that the integrand is a periodic function
in $\varphi$ with period $2\pi$, we can perform a final change of
variables $\theta=\varphi-i\varphi_0$ to find
\widetext
\beq
\label{eq:A4}
A(\tau,X)=\theta[(b X)^2 - \tau^2]{1\ov 2b}\int_0^{2\pi}
{d\theta\ov 2\pi}e^{a\sqrt{X^2-(\tau/b)^2}\cos\theta}
={1\ov 2b}\theta[(b X)^2 - \tau^2]
I_0\left({{a\ov b}\sqrt{(b X)^2-\tau^2}}\right),
\eeq
where $I_0$ is a Bessel function of imaginary argument.

Using $a=\Delta_{\xi}, b=\lambda/\pi, \tau=v_{\rm n}X-t$ and differentiating
Eq.~(\ref{eq:A4}) with respect to $\tau$ and $X$ gives the following results 
\beqarr
\ii {d\omega\ov 2\pi}e^{i\omega(v_{\rm
n}X-t)}{\sinh(\Lambda(\omega)X)\ov \Lambda(\omega)}&=&
{1\ov 2(\lambda/\pi)}\theta(z)
I_0\left({{\Delta_{\xi}\ov\lambda/\pi}\sqrt{z}}\right),\label{eq:sh/L}\\
\ii {d\omega\ov 2\pi}e^{i\omega(v_{\rm n}X-t)}\cosh(\Lambda(\omega)X)&=&
{1\ov 2}\sgn(t)\left[{\delta(x/v_1-t)+\delta(x/v_2-t)}\right]
+ {\Delta_{\xi}\ov 2v_1v_2}\theta(z)\nonumber\\
&&{}\times{x\ov \sqrt{z}}
I_1\left({{\Delta_{\xi}\ov\lambda/\pi}\sqrt{z}}\right),\label{eq:ch}\\
\ii {d\omega\ov 2\pi}e^{i\omega(v_{\rm n}X-t)}
i\omega{\sinh(\Lambda(\omega)X)\ov\Lambda(\omega)}&=&
-{1\ov 2\lambda/\pi}\sgn(t)\left[{\delta(x/v_1-t)-\delta(x/v_2-t)}\right]
\nonumber\\
&&{}-{\Delta_{\xi}\ov 2(\lambda/\pi)^2}\theta(z)
{(v_{\rm n}x/v_1v_2-t)\ov \sqrt{z}}
I_1\left({{\Delta_{\xi}\ov\lambda/\pi}\sqrt{z}}\right)\!,
\label{eq:omega_sh/L}
\eeqarr
where $z\equiv (t-x/v_1)(x/v_2-t)$.

Finally, if we analytically continue the above results to the case of
$\Delta_{\xi}=i\sqrt{v_1v_2}\Gamma$, where $\Gamma$ is real, and
write $\kappa(\omega)=\sqrt{v_1v_2\Gamma^2+(\lambda/\pi)^2\omega^2}$,
we find
\beqarr
\ii {d\omega\ov 2\pi}e^{i\omega(v_{\rm
n}X-t)}{\sin(\kappa(\omega)X)\ov \kappa(\omega)}&=&
{1\ov 2(\lambda/\pi)}\theta(z)
J_0\left({{\Gamma\ov\lambda/\pi}\sqrt{v_1v_2z}}\right),\label{eq:sin/k}\\
\ii {d\omega\ov 2\pi}e^{i\omega(v_{\rm n}X-t)}\cos(\kappa(\omega)X)&=&
{1\ov 2}\sgn(t)\left[{\delta(x/v_1-t)+\delta(x/v_2-t)}\right]
-{\Gamma\ov 2}\theta(z)\nonumber\\
&&{}\times{x\ov \sqrt{v_1v_2z}}
J_1\left({{\Gamma\ov\lambda/\pi}\sqrt{v_1v_2z}}\right),\label{eq:cos}\\
\ii {d\omega\ov 2\pi}e^{i\omega(v_{\rm n}X-t)}
i\omega{\sin(\kappa(\omega)X)\ov\kappa(\omega)}&=&
-{1\ov 2\lambda/\pi}\sgn(t)\left[{\delta(x/v_1-t)-\delta(x/v_2-t)}\right]
\nonumber\\
&&{}+{\Gamma\theta(z)\ov 2(\lambda/\pi)^2}
{(v_{\rm n}x-v_1v_2t)\ov \sqrt{v_1v_2z}}
J_1\!\left({\Gamma\sqrt{v_1v_2z}\ov\lambda/\pi}\right)\!.
\label{eq:omega_sin/k}
\eeqarr
\narrowtext


\begin{thebibliography}{1}
\bibitem{endnote1}
If one includes the possibility of
constrictions one can also probe the dimensions of quasiparticle
operators.
\end{thebibliography}
\end{document}